\documentclass[fleqn,usenatbib]{mnras}

\usepackage{newtxtext,newtxmath}

\usepackage[T1]{fontenc}

\usepackage{array}
\newcolumntype{L}[1]{>{\raggedright\let\newline\\\arraybackslash\hspace{0pt}}m{#1}}
\newcolumntype{C}[1]{>{\centering\let\newline\\\arraybackslash\hspace{0pt}}m{#1}}
\newcolumntype{R}[1]{>{\raggedleft\let\newline\\\arraybackslash\hspace{0pt}}m{#1}}

\DeclareRobustCommand{\VAN}[3]{#2}
\let\VANthebibliography\thebibliography
\def\thebibliography{\DeclareRobustCommand{\VAN}[3]{##3}\VANthebibliography}

\usepackage{graphicx}	
\usepackage{amsmath}	
\usepackage{csvsimple,booktabs,array,siunitx}
\usepackage{tabularx}
\usepackage{adjustbox} 
\usepackage[figuresright]{rotating}
\usepackage{caption}
\newcommand{\noopsort}[2]{#2}
\usepackage{hyperref}

\title[GRGs in the Bo\"otes deep field]{Giant radio galaxies in the LoTSS Bo\"otes deep field}

\author[M. Simonte et al.]{
M. Simonte,$^{1}$\thanks{E-mail: marco.simonte@hs.uni-hamburg.de}
H. Andernach,$^{2}$
M. Br\"uggen,$^{1}$
D.~J. Schwarz,$^{3}$
I. Prandoni,$^{4}$
A.~G. Willis$^{5}$
\\
$^{1}$ Hamburger Sternwarte, University of Hamburg, Gojenbergsweg 112, 21029 Hamburg, Germany\\
$^{2}$Departmento de Astronomía, DCNE, Universidad de Guanajuato, Callejón de Jalisco s/n, Guanajuato, C.P. 36023, GTO, Mexico; heinz@ugto.mx\\
$^{3}$ Fakult\"at f\"ur Physik, Universit\"at Bielefeld, Postfach 100131, 33501 Bielefeld, Germany\\
$^{4}$ INAF – Istituto di Radioastronomia, via Gobetti 101, I-40129 Bologna, Italy\\
$^{5}$ Independent Researcher, 310 Yorkton Avenue, Penticton, BC V2A 6Z8, Canada
}

\date{Accepted XXX. Received YYY; in original form ZZZ}

\pubyear{2022}

\begin{document}
\label{firstpage}
\pagerange{\pageref{firstpage}--\pageref{lastpage}}
\maketitle

\begin{abstract}
Giant radio galaxies (GRGs) are radio galaxies that have projected linear extents of more than 700 kpc or 1 Mpc, depending on definition. We have carried out a careful visual inspection in search of GRGs of 
the Bo\"otes LOFAR Deep Field (BLDF) image at 150 MHz. We identified 74 GRGs with a projected size larger than 0.7 Mpc of which 38 are larger than 1 Mpc. The resulting GRG sky density is about 2.8 (1.43) GRGs per square degree for GRGs with linear size larger than 0.7 (1) Mpc. We studied their radio properties and the accretion state of the host galaxies using deep optical and infrared survey data and determined flux densities for these GRGs from available survey images at both 54 MHz and 1.4 GHz to obtain integrated radio spectral indices. We show the location of the GRGs in the P-D diagram. The accretion mode onto the central black holes of the GRG hosts is radiatively inefficient suggesting that the central engines are not undergoing massive accretion at the time of the emission. Interestingly, 14 out of 35 GRGs for which optical spectra are available show a moderate star formation rate (10-100 $\rm M_{\odot}~yr^{-1}$). Based on the number density of optical galaxies taken from the DESI DR9 photometric redshift catalogue, we found no significant differences between the environments of GRGs and other radio galaxies, at least for redshift up to $z=0.7$.
\end{abstract}




\section{Introduction}
 \label{Intro}


Giant radio galaxies (GRGs) are radio galaxies that have linear extents of more than 700 kpc or 1 Mpc, depending on definition \citep{Willis1974, Schoenmakers2000, Kuzmicz2012, Kuzmicz2018}.   The total number of GRGs known to date is relatively small, even though over the past 20 years their number has increased substantially \citep{Ishwara-Chandra1999, Schoenmakers2000, Lara2000, Machalski2001, Saripalli2005, Machalski2006,  Jamrozy2008, Kuzmicz2012, Dabhade2017, Kuzmicz2018, Dabhade2020, SaganI, SaganII, Kuzmicz2021, Andernach2021, SAGANIII}. Their origin and the cause for their huge sizes are still not understood.

In particular, it is unclear whether their environments or their host properties are responsible for the large extent of GRGs \citep[e.g,][]{Subrahmanyan2008, Safouris2009, Kuzmicz2019}. The environment of GRGs might play a key role, affecting the accretion mode onto the central black hole and suppressing the expansion of the radio lobes in the case the GRG resides in a dense region. This would in turn lead to larger sources in more isolated galaxies \citep{Dabhade2020, Andernach2021}. Alternatively, the large size of these galaxies might be the result of a long-term evolution of normal radio galaxies \citep{Kaiser1997, Kaiser1997a}.


The advent of large-area radio surveys, such as the Faint Images of the Radio Sky at Twenty-cm \citep[FIRST,][]{FIRST1995}, Westerbork Northern Sky Survey \citep[WENSS,][]{WENSS1997}, NRAO VLA Sky Survey \citep[NVSS,][]{NRAO1998} and the Sydney University Molonglo Sky Survey \citep[SUMSS,][]{SUMSS2003} ushered in a new era during which systematic studies of GRGs were carried out.  Early studies \citep{Subrahmanyan1996, Saripalli1996, Mack1998} investigated the radio properties of GRGs (such as the axial ratio) finding similarities with those of smaller radio galaxies, indicative of self-similarity in the evolution of such sources \citep{Kaiser1997}. These authors also found that GRGs preferentially reside in low-density environments. Furthermore, \citet{Schoenmakers2000} and \citet{Lara2000} found their radio spectra to be steep ($0.8 < \alpha < 1.2$) in the lobes of GRGs, corresponding to ages of 10 -100 Myr. More recently, \citet{Jamrozy2008}, using data from the Giant Metrewave Radio Telescope \citep[GMRT,][]{GMRT}, estimated spectral ages in the range 5 - 40 Myr, observing a steeper spectral slope at high luminosities and high redshifts. They also found steeper spectra for larger GRGs. \citet{Safouris2009} and \citet{Subrahmanyan2008} found that asymmetries in the radio morphology of GRGs may be driven by the inhomogeneities of the surrounding medium. In particular, both studies suggest that the jets of GRGs expand into regions that are relatively sparsely populated by galaxies.  \citet{Machalski2008} argued that a combination of a low-density environment and jet speeds of about 0.1c - 0.2c has led to the formation of J1420-0545 which is now the second largest GRG known \citep{Alcyneous2022} with a linear size of 4.69 Mpc.

 Subsequent work tried to model the radio data to infer the age, the spectral index and the environment of GRGs \citep[e.g.][]{Machalski2007, Machalski2009} confirming a spectral age between 10-100 Myr. In particular, \citet{Machalski2011}, using the DYNAGE algorithm \citep{Machalski2007}, found a relation between the difference in value of the exponent describing the external gas density profile ($  \rho_{\rm IGM} \propto r^{-\beta}$) for the opposite lobes and the ratio of their volume, suggesting the non-uniformity of the environment of the GRGs. 
 
 More recent work on GRGs boosted the study providing larger samples \citep{Dabhade2017, Kuzmicz2021, Dabhade2020, Andernach2021}, but a clear theory of the evolution of such radio sources is still missing. In those studies a few per cent of GRGs were found to reside in galaxy clusters, challenging the idea that they are typically associated with underdense environments. According to these studies, GRGs have radio powers similar to normal (i.e. smaller) radio galaxies, in agreement with the evolutionary theory \citep{Kaiser1997}. 
Altogether these analyses revealed a prevalence of FRII \citep{FR} morphology among GRGs, altough other morphologies have been detected, such as double-double radio galaxies \citep[DDRGs,][]{DDRG2000}, Hybrid Morphology Radio Sources \citep[HyMoRS,][]{Gopal-Krishna2002}, which are radio galaxies with a FRI type radio lobe on one side of the active nucleus and a FR II type lobe on the opposite side, and Wide Angle Tailed radio galaxies \citep[WAT,][]{Owen1976}.

Infrared and optical surveys can be exploited to identify GRG host galaxies and investigate the galaxy distribution around them \citep[e.g.,][]{Lan2021}. Among these surveys are the Panoramic Survey Telescope and Rapid Response System (Pan-STARRS; \citealt{panstarss}), Sloan Digital Sky Survey (SDSS; \citealt{SDSS1, SDSS2, SDSS3}), Dark Energy Spectroscopic Instrument survey \citep[DESI;][]{Legacy2019, Zhou2021} and the Wide-field Infrared survey Explorer (WISE; \citealt{Cutri2013, Marocco2020}). A large amount of data have been used to infer information on the optical counterpart and the distribution of galaxies around GRGs \citep[e.g.,][]{Lan2021}. For instance, the SAGAN project \citep[see][for details]{SaganI} aims to create a catalogue of all GRGs published to date including the properties of the host galaxies.

 There is ample theoretical work on the evolution of radio galaxies \citep[e.g.][]{Burns1991, Kaiser1997a, Machalski2021}. Recent and refined studies investigated the propagation of powerful jets and the impact with the surrounding environment (interstellar or intergalactic medium) at distances of tens to hundreds of kiloparsecs \citep{Mignone2010, English2016, Hardcastle2018, Perucho2021, Yates2022}. However, the computational limitations make the study of the long-term simulations with an adequate resolution challenging.

The LOw-Frequency ARray \citep[LOFAR][]{LOFAR2013} with its relatively high resolution and sensititvity to very low surface brightness sources heralds a new era in the study of very large and high redshift radio galaxies. In this paper we searched for GRGs in the Bo\"otes LOFAR Deep Field (BLDF) image at 150 MHz \citep{Tasse2021, Kondapally2021}. The resulting sample contains 74, mostly newly detected, GRGs. By cross-matching with other radio data, we obtained the radio spectral information and by cross-matching with optical and infrared data, we inferred the properties of the host galaxies as well as environments of the GRGs.
 
 

The outline of this paper is as follows: we describe the analysis of the radio data, as well as the infrared data in Sec.~\ref{sec:observations}. In Sec.~\ref{sec:results} we present the results of our analysis and compare them to previous work. Finally, we draw our conclusions in Sec.~\ref{sec:conclusion}. 
Throughout this work we adopt a flat $\rm \Lambda CDM$ cosmology with  \textit{$H_0$} = 70 $\rm ~ km ~ s^{-1} ~ Mpc^{-1}, ~ \Omega_m = 0.3, ~ \Omega_{\Lambda} = 0.7$ and a radio source spectral index $\alpha$ defined as $S_{\nu} \propto \nu^{-\alpha}$.

\section{Observations and multi-frequency data}
\label{sec:observations}

The LOFAR Two-metre Sky Survey (LoTSS; \citealt{LoTSS1}, \citeyear{LoTSS2}, \citeyear{DR2}) is performed using high-band antenna (HBA) observations of the northern sky in the 120-168 MHz band and images at both 20$ \rm "$ and 6$\rm "$ resolution are available. The sensitivity of the latter image is about 1$\sigma= 80 \rm ~ \mu Jy ~ beam^{-1}$. LoTSS provides the astrometric precision that is required for multi-wavelength cross-matching.  As part of this survey, dedicated deeper observations have been carried out of the Bo\"otes, the Lockman Hole \citep{Tasse2021} and the ELAIS-N1 fields \citep{Sabater2021}. The long integration time ($>$ 80 hours) combined with a sensitivity to a wide range of angular scales, makes these deep fields ideal for a search for the faintest and
most distant GRGs. We have carried out a systematic search of GRGs in the Bo\"otes LOFAR Deep Field (centered on 14h32m03.0s +34$^\circ$16$\rm '$33$\rm ''$ J2000) which covers an area of 26.5 deg$^2$ and with a noise level of $\sim$30 $\rm \mu Jy ~ beam^{-1}$ in the inner 3 deg$^2$. A dedicated search for GRGs in the other deep fields will be part of a future study.

The inspection of LOFAR images was carried out by eye. The largest angular size (\textit{LAS}) in arcmin, was measured on the full resolution 6$\rm "$ image. \citet{Kuzmicz2021} showed  that an estimate based on 3-$\sigma$ contours would make the source appear larger by at least one beam size for sources of FR-II type with bright compact hotspots at the outer edges of their lobes. Therefore, while for FRI we used the 3-$\sigma$ contours to measure the \textit{LAS}, for FRII type GRGs we measured the distance between the two opposite hotspots, identified on VLA Sky Survey images \citep[VLASS,][]{vlass2020} if available. On the other hand, we believe that an angular size measurement based on the 3-$\sigma$ contours underestimates the real size of those sources of FR\,I type, or those which have no clear terminal hotspots in their lobes, some of which clearly extend beyond the 3-$\sigma$ contour. For this reason, for the faintest GRGs we measure the angular size in a straight line from one end of the source to the opposite one, carefully avoiding to push our measurement in regions where the radio emission of the sources is not very clear. The same method has been used to measure the angular size of bent sources; any measure different from this would imply assumptions on the 3D structure of the source which we can only guess. We found 74 GRGs with a linear extent larger than 0.7 Mpc (which we refer to as the "Bo\"otes LOFAR Deep Field" or BLDF-GRG sample). The only GRG reported in a GRG catalogue is J1427+3625 in \citet{Kuzmicz2021}, while J1430+3519 was studied in the radio and X-ray band by \citet{Masini2021}. 16 GRGs were listed in other catalogues of radio galaxies but they were published without their linear size \citep[][]{Williams2013, VanWeeren2014, Coppejans2015, Williams2016, Mingo2022}. Among our sample, 38 GRGs have a linear extent larger than 1 Mpc. The resulting sky density is about 2.8 GRGs deg$^{-2}$ for GRGs with linear size larger than 0.7~Mpc 
and 1.43 deg$^{-2}$ for those larger than 1~Mpc. This result is in agreement with recent work by \citet{Delhaize2021}, who found two GRGs in the 1 square degree COSMOS field, and higher than in \citet{Brueggen2021} who found a sky density of radio galaxies with largest linear size $>0.7$ Mpc of 1.7 deg$^{-2}$ in ASKAP observation of the Abell 3395-Abell 3391. This field has similar noise level, but lower angular resolution (10$"$) than the BLDF. 

 The host galaxies were identified using optical and infrared surveys: SDSS \citep{SDSS1}, WISE \citep{WISE2010} and its multiple catalogues such as AllWISE \citep{Cutri2013}, unWISE \citep{unWISE2019} and CatWISE \citep{Marocco2020}, DESI \citep{Legacy2019, Zhou2021} and Pan-STARRS \citep{panstarss}. Once the host galaxy has been identified  we looked for available redshifts (either spectroscopic or photometric) in multiple catalogues such as \citet{El2009}, \citet{Chung2014}, \citet{Brescia2014}, \citet{lamost}, \citet{Bilicki2016}, \citet{Zou2019}, \citet{Duncan2021}. If various redshifts were available for a single source we computed their mean. For spectroscopic redshifts we do not report errors since they are generally more accurate (typical errors are usually around 0.00015) than the precision we can achieve on the linear sizes and luminosities  given our errors in measuring angular size and total flux. The identification of the host galaxy in widely separated radio components without an obvious core emission between them is challenging. We first checked whether any of the outer components have a convincing host by itself. In such cases, they were recorded as separate extended radio galaxies. If a convincing optical galaxy between the external radio components is detected in neither optical nor infrared surveys, we discarded the source as genuine GRG. On the other hand, we take the presence of a host galaxy with AGN colour \citep[e.g.,][]{Mateos2012, Assef2013} near the geometrical centre of the radio galaxy as a strong sign of the genuineness of the GRG, even if that host does not show a radio core. Furthermore, the presence of a radio bridge or radio emission elongated along the suspected radio axis sometimes indicates the structures to be connected. In case of uncertainties about the likely host we chose the brighter or lower redshift host. In such sources, the largest linear size (\textit{LLS}) derived from the \textit{LAS} and redshift should serve as a lower limit. Three GRGs, namely J1421+3521, J1423+3340 and J1434+3214C, do not have a redshift estimate in any of the mentioned catalogues and they are not detected in DESI DR9. The photometric redshift provided by the DESI DR9 photometric redshift catalogue (accessible via the NOIRlab portal\footnote{\href{https://astroarchive.noirlab.edu/portal/search/}{https://astroarchive.noirlab.edu/portal/search/}}) for the faintest objects is around 1.3. For this reason, we assumed that the GRG hosts are located at redshift \textit{z}=1.1-1.5. However, the exact value of the redshift is quite irrelevant for
 converting \textit{LAS} to \textit{LLS} since the \textit{LAS-z} curve is practically flat in the relevant redshift range (see the \textit{LAS-z} diagram in sect. 3.2), so it does not affect the \textit{LLS} significantly. 


 We used the full resolution 6$\rm "$ images to measure the total radio flux at 150 MHz. We prepared cutouts of 0.5 degrees on a side, centered on the GRG host and integrated the flux of GRGs taking into account only those pixels whose intensity is larger than $3\sigma_{\rm rms}$, where $\sigma_{\rm rms}$ is the noise level in the source neighbourhood. This noise level varies with the distance from the centre of the BLDF image, but it is rather constant within each cutout.  
 The flux error was calculated as $\sqrt{\sigma_{\rm rms}^2 + \sigma_{\rm cal}^2}$, where $\sigma_{\rm cal}$ is the uncertainty on the calibration of the flux scale which is assumed to be 10\% of the total flux \citep{Sabater2021}. We measured the noise of the cutout, $\sigma_{\rm rms}$, via an iterative method. For each iteration, we calculated the rms, we removed those pixels with an intensity larger than 5 times the rms and, consequently, we re-measured the rms. The convergency criterion is dictated by the difference between two consecutive rms measurement; the treshold was set equal to 1\%. Thus, we multiply the noise measurement by the square root of the area of flux integration, measured in units of beam areas. 
We noticed that a 3-$\sigma$ clipping applied to the faintest GRGs clearly underestimates the total flux of the source as some genuine emission coming from the bridge is not included. Thus, we integrated the flux of such sources, marked by an asterisk appended to the flux measure in Table~\ref{tab:list}, considering all the pixels belonging to the region of the radio emission.
 Radio powers at 150 MHz for GRGs were calculated following \citet{Donoso2009}:

\begin{equation}
      P_{150} = 4 \pi D_L^2 S_{150} (1+z)^{\alpha - 1} ,
      \label{eq:power_estimation}
\end{equation}

where $ D_L$ is the luminosity distance, $ S_{150}$ is the measured radio flux density at  150 MHz, $(1+z)^{\alpha - 1}$ is the standard k-correction used in radio astronomy and $\alpha$ is the radio spectral index for which we adopted a typical value of 0.7. The median radio power is $ P_{\rm med} = 6.61\times10^{25} \rm ~ W ~ Hz^{-1}$ and the GRGs span a wide range of redshift, up to $z \sim 2.8$, with a median redshift of $z_{\rm med} = 0.80$. \citet{Kuzmicz2018} list 10 GRGs with $10^{23} < P_{1400} < 10^{24}$ at low redshift, while \citet{Dabhade2017} and \citet{Dabhade2020} found GRGs down to  Log($P_{150}$) $\sim 24$. The median r-band magnitude of the host galaxies in our BLDF-GRG sample is 21.68, compared to 20.7 for the 181 GRGs that \citet{Andernach2021} found by inspecting the 888-MHz Rapid ASKAP Continuum Survey \citep[RACS,][]{McConnell2020}, showing the Bo\"otes LOFAR deep field to allow finding fainter and more distant radio galaxies. A quantitative comparison with some of the most recent and largest samples of GRGs is shown in Table~\ref{tab:comparison}. The columns are: (1) reference and survey used, (2) the observing frequency, (3) number of GRGs in the sample, (4) minimum and median flux of the sample at the observing frequency, (5) minimum and median decimal logarithm of the power of the sample at 150 MHz calculated assuming a spectral index $\alpha$=0.7, (6) median of the redshift of the host galaxies. It is worthwhile to notice that the median flux of the BLDF-GRGs is the lowest among the reported samples, while the median redshift is the second largest, only slightly surpassed by \citet{Kuzmicz2021} who carried out a dedicated search for extended radio galaxies with \textit{LLS}$>$1 Mpc from spectroscopic QSOs from SDSS DR14Q. 

The list of our sources is provided in Table~\ref{tab:list}. A "C" appended to the GRG name indicates that it is a candidate, meaning that either the host itself, or its redshift, or its \textit{LAS} are uncertain. Fig.~\ref{fig:GRG_images} shows the cutouts of the BLDF image around the GRGs. The size of these cutouts is proportional to the \textit{LAS} of the GRG and it is smaller than the size of the cutouts used for the flux integration. The cyan circle identifies the position of the galaxy host. Notes on individual GRGs are reported in the appendix. 

 
 \begin{center}
 \begin{table}
    \centering
    \begin{tabular}{p{1.9cm}p{0.5cm}p{0.5cm}p{1.3cm}p{1.1cm}p{0.4cm}}
    \hline
      (1) & (2) & (3) & (4) & (5) & (6) \\
      Survey,Reference & Freq. & N of & $S_{\rm min,med}$ & $P_{\rm min,med}^{150}$ & $z_{\rm med}$ \\ 
      & GHz & GRGs  & mJy & W/Hz &  \\ \hline
      Literature$^1$ & 0.8/1.4 & 349 & 5.20,164 & 23.7,26.2 & 0.24 \\ 
      NVSS$^2$ & 1.4 & 25 & 28.0,95.0 & 24.7, 25.8 & 0.22 \\
      LOTSS$^3$ & 0.15 & 239 & 2.00,218 & 24.0,26.1 & 0.53 \\
      NVSS$^4$ & 1.4 & 161 & 3.00,209 & 24.4,25.3 & 0.23 \\
      NVSS,SDSS$^5$ & 1.4 & 76 & ----,---- & 26.0,26.7 & 0.82 \\
      RACS$^6$ & 0.888 & 181 & 5.00, 40.0 & 24.0,26.4 & 0.66 \\
      BLDF, this work & 0.15 & 74 & 3.6, 29.2 & 24.5,25.8 & 0.80 \\
 \hline
    \end{tabular}
    \caption{Comparison between our BLDF-GRG and previous samples. References: 1-\citet{Kuzmicz2018}, 2-\citet{Dabhade2017}, 3-\citet{Dabhade2020}, 4-\citet{SaganI}, 5-\citet{Kuzmicz2021}, 6-\citet{Andernach2021}. References 5 and 6 are compilations of GRGs with \textit{LLS} larger than 1 Mpc, while all the others include GRGs with \textit{LLS} $>$ 0.7 Mpc.}
    \label{tab:comparison}
\end{table}
\end{center}

\subsection{LoLSS and NVSS data}
\label{sec:multifreq}

Data from radio surveys carried out at different frequencies allow an analysis of the spectra of the radio sources. However, the low surface brightness of GRGs makes their detection challenging, especially at higher frequencies. The only deep observation in the Bo\"otes field at higher frequencies is presented in \citet{devries2002} although the image is not available. Moreover, the APERture Tile In Focus (Apertif) survey \citep{vancappellen2022} is not as complete as NVSS in this region. For this reason, we looked for the BLDF-GRGs in the NVSS \citep{NRAO1998} which covers the sky north of -40 deg declination at 1.4 GHz. The NVSS has an angular resolution of 45$\rm "$ and is very sensitive to sources with low surface brightness. The rms noise level is 0.45 $\rm mJy ~ beam^{-1}$ (Stokes I). Finally, the LOFAR LBA Sky Survey \citep[LoLLS,][]{LoLSS} covers the very low-frequency ($< 100 \rm ~ MHz$) regime. The LoLLS Bo\"otes deep field was observed with the Low Band Antennas (LBA) at  34-75 MHz for 56 hours \citep{Williams2021}. The integration time makes this observation the first sub-mJy survey below 100 MHz, with an rms noise of 0.7 $\rm mJy ~ beam^{-1}$. The resulting image has an angular resolution of 15$\rm "$.

\subsection{Infrared data}

\begin{figure}
        \centering
        \includegraphics[width= 0.5\textwidth]{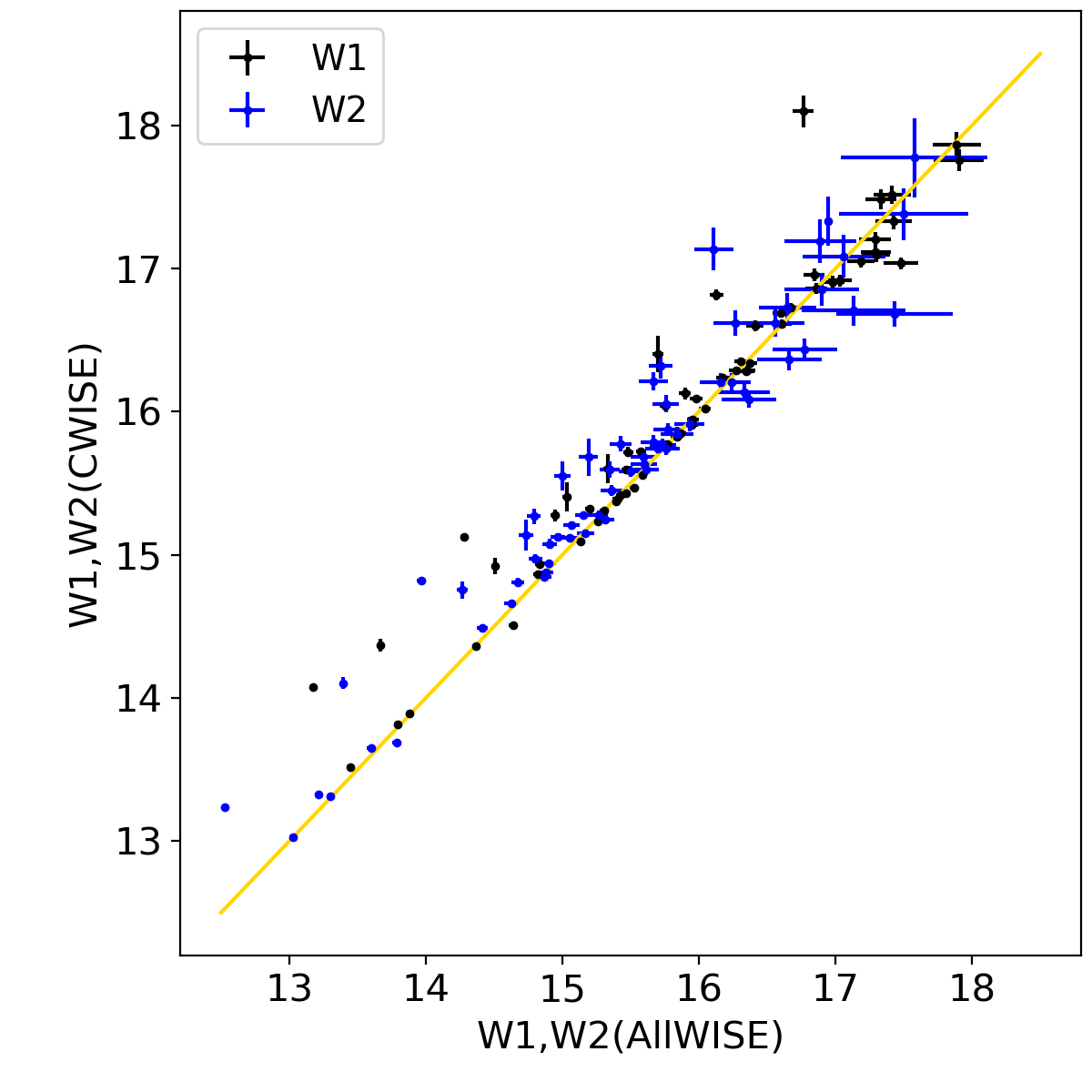}
        \caption{Comparison between the magnitudes in the W1 (black) and W2 (blue) bands in the AllWISE and CWISE catalogue. The yellow line marks the one-to-one relation.}
        \label{fig:magnitude_test}
  \end{figure}

We cross-matched the GRG hosts with infrared data from the WISE survey in order to determine the accretion mode of the central engine. 
The Wide-field Infrared Survey \citep{WISE2010} is an all-sky survey conducted in four spectral bands: W1 ($3.4 \mu$m), W2 ($4.2 \mu$m), W3($12 \mu$m), W4 ($22 \mu$m) with angular resolution 6.1, 6.4, 6.5, 12$\rm "$, respectively. AllWISE \citep{Cutri2013} is the resulting catalogue of the combination of WISE and NEOWISE \citep{Mainzer2011} surveys. The AllWISE source catalogue contains accurate positions, proper motion measurements, four-band fluxes and flux variability statistics.The CatWISE2020 (CWISE) catalogue contains objects selected from WISE and NEOWISE survey data at 3.4 and 4.6 $\mu$m (W1 and W2) and it is the most extensive dataset of the full mid-infrared sky. We found that the CatWISE catalogue often (in about 25\% of cases) provides multiple matches for a single position with a distance between the matches less than the  angular resolution of the telescope.  Moreover, often the source with the brighter magnitude appears with lower S/N ratio, opposite to expectation. Furthermore, the catalogue does not have data at 12 and 22 $\mu$m that are used in the WISE color-color diagram. Comparing the AllWISE and CatWISE magnitudes, we found that the W1 and W2 magnitudes in both catalogues are consistent with each other (see Fig.~\ref{fig:magnitude_test}). We decided to use the AllWISE data for our infrared analysis. The CWISE catalogue contains fainter sources than AllWISE and eight of the GRG hosts are only detected in CWISE. Hence, we did not consider these for the infrared analysis. We noticed that two types of magnitudes are listed, namely "Wipro" and "Wimag", where i=1,2,3,4 refers to the WISE band. The former is the magnitude measured with profile-fitting photometry, while the latter is the magnitude measured in an 8.25$\rm "$ radius circular aperture centered on the source position \citep[see][for more details]{Cutri2013} \footnote{ \href{https://wise2.ipac.caltech.edu/docs/release/allwise/expsup/sec2\_1a.html}{https://wise2.ipac.caltech.edu/docs/release/allwise/expsup/sec2\_1a.html}}. The NASA/IPAC Infrared Science Archive \footnote{ \href{https://wise2.ipac.caltech.edu/docs/release/allsky/}{https://wise2.ipac.caltech.edu/docs/release/allsky/}} provides the WISE magnitudes "Wipro" and the associated errors in the four bands. 
  
 \section{Results}
 \label{sec:results}

 \subsection{Distribution of the largest linear sizes}

 The distribution of the \textit{LLS} of the (giant) radio galaxies can give clues as to whether GRGs are just extreme cases of the general population of radio galaxies (RGs), or constitute an independent class of sources.
 
 There are several distributions that describe either the extreme values of another underlying distribution, such as the distribution of the largest size for a given sample of RGs, or distributions for random objects above a given threshold, such as the exponential, or the Pareto distribution \citep[see e.g.][for more details]{Coles_2001}. GRGs are part of the latter type of distributions as they are RGs whose size exceeds 0.7 Mpc. As a consequence, we restrict our analysis to exponential (as a special case) and generalised Pareto \citep{Coles_2001} distributions. The latter is commonly used to estimate the probability of exceedances over a high threshold \citep[see][for applications]{Zaninetti2008, Bouillot2015, Aschwanden2015}. The generalised version of such a distribution is:
\begin{equation}
    f_P(x) = \frac{1}{\sigma}\left (1+c\frac{x-\mu}{\sigma}\right )^{-1-\frac{1}{c}},
    \label{eq:pareto}
\end{equation}
where $c$ is the shape parameter. For $c > 0$ the support of the
distribution is $x \geq \mu$, while for $c < 0$ it is limited to 
$\mu \leq x \leq \mu - \sigma/c$. The exponential distribution is 
a special case of the generalised Pareto distribution obtained by 
taking the limit $c \rightarrow 0$:
\begin{equation}
    f_E(x) = \frac{1}{\sigma}e^{-\frac{x-\mu}{\sigma}}, \quad x \geq \mu.
    \label{eq:expo}
\end{equation}
In Eqs.~(\ref{eq:pareto}) and (\ref{eq:expo}), 
$\mu$ and $\sigma > 0$ are the 
location (the threshold value) and scale parameters, respectively.
 
 Recently, Oei et al. (in prep.) carried out a detailed analysis of the distribution of the \textit{LLS} using about 500 GRGs. Assuming a Pareto distribution, the authors found a slope of $-4$, thus $c=1/3$ (see Eq.~\ref{eq:pareto}). These authors mostly focused on local GRGs ($z < 0.2$) with an angular size larger than 5$\rm '$, while \citet{Andernach2021} observed a similar slope in the cumulative size distribution of a compilation of GRGs found by visual inspection of sources 
 in the RACS survey \citep{McConnell2020} complete down to an angular size of $\sim2 \rm '$ and independent of redshift, which resulted in 178 new GRGs larger than 1 Mpc.

 In this work, we carried out a simple statistical analysis on the \textit{LLS} distribution of our GRGs, using the complete sample of extended RGs provided by \citet{Miraghaei2017} (MB17 hereafter). This sample provides measurements of the largest linear sizes for 1329 extended RGs (we revised the measurement of the \textit{LAS} at 1.4 GHz of some sources and consequently re-calculated the linear size by using the listed redshifts) with hosts selected from SDSS. The flux density threshold of this sample is 40 mJy, which corresponds to roughly 200 mJy at LoTSS frequencies for $\alpha = 0.7$. The catalogue shows a lack of RGs with linear extents smaller than 50 kpc and complete catalogues for the smallest RGs are missing in literature \citep{Capetti2017, Capetti2020}. For this reason, we did not include the RGs with \textit{LLS} $<$ 50 kpc in our analysis.
 
\begin{figure}
        \centering
        \includegraphics[width= 0.5\textwidth]{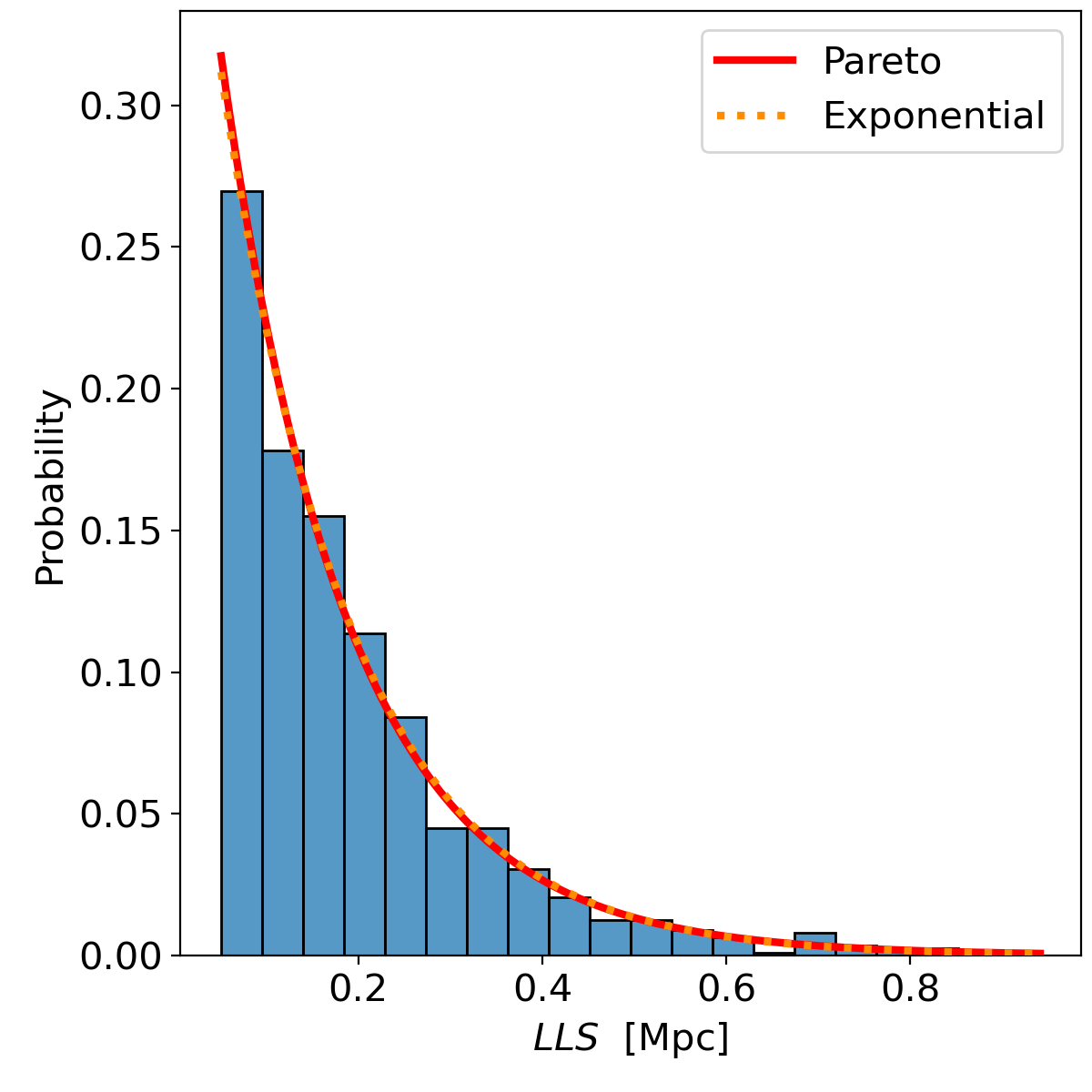}\quad \includegraphics[width= 0.5\textwidth]{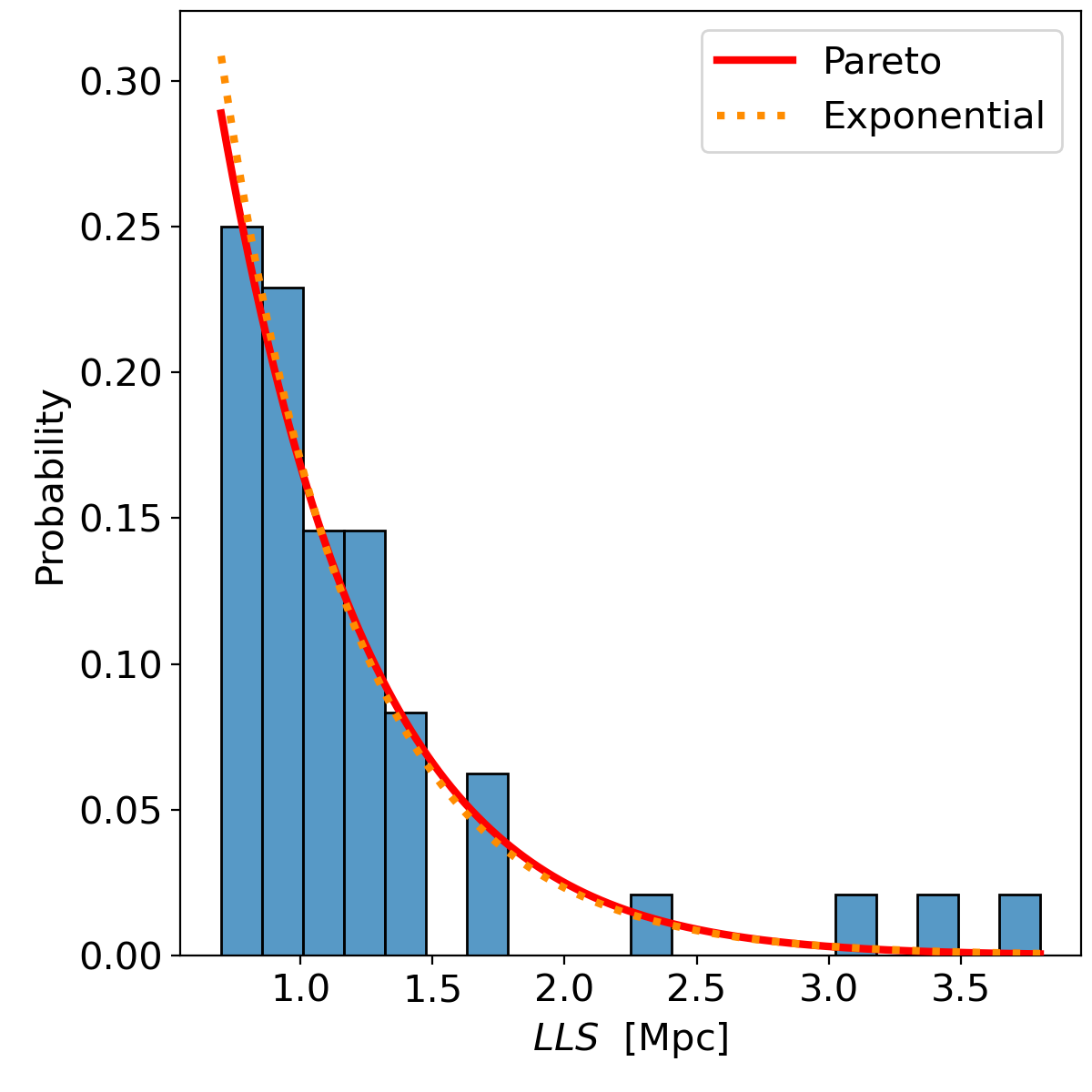}
        \caption{Top panel: Distribution of the largest linear sizes of a sample of RGs provided by \citet{Miraghaei2017}. 
        Lower panel: Distribution of the largest linear sizes in the BLDF-GRG sample.
        The solid red and dotted orange lines are the best fit of the Pareto and exponential distribution respectively. The $y$-axis represents the probability that a (G)RG has a certain \textit{LLS}.}
        \label{fig:lls_test}
\end{figure}

 We used maximum likelihood estimation to fit the two distributions to the \textit{LLS} of our GRGs and the RGs listed in MB17.
 The parameters of the fit are the location and scale parameters; the shape parameter, $c$, is considered only in the generalised Pareto distributions. In this analysis, the location parameter is the minimum size of the RGs of the two samples (50 kpc and 700 kpc for the MB17 and BLDF sample respectively). Thus, we performed a Kolmogorov-Smirnov (KS) test between our samples and randomly generated data coming from the resulting best-fit distributions for 1000 times, averaging the p-values obtained from each iteration. Such a test is commonly used to decide whether two samples derive from the same population; the closer the p-value is to 0 the more confident we are in rejecting that the samples were drawn from the exponential or Pareto distribution. The results are summarised in Table~\ref{tab:lls_test} and we show the fit in Fig.~\ref{fig:lls_test}. The location and scale parameters are given in units of Mpc .
 
 For the  MB17 sample we found $c=0.02 \pm 0.01$ when fitting the generalized Pareto distribution, which indicates that the distribution of the \textit{LLS} is very close to exponential. The p-value resulting from the KS test between the MB17 sample and the exponential distribution is $0.52$, confirming that the exponential distribution provides an acceptable description of the data.
 Similarly, both distributions are reasonable fits to the sample of the \textit{LLS} of the GRGs. The rather small value of the shape parameter, $c=0.22 \pm 0.12$, suggests that also the \textit{LLS} of the GRGs follows a fairly steep Pareto distribution. It is also consistent with an exponential distribution with the same scale parameter. 
 
 However, the scale parameters of the RG and GRG samples are different. The scale parameter of the MB17 sample, which is 140 kpc, is well above the cutoff suggesting that the typical size of the extended RGs is larger than 50 kpc. On the other hand, the scale for GRGs is 0.52 Mpc which is below the cutoff, showing that GRGs are indeed larger than the intrinsic scale of the distribution. These results indicate that, even though we are not probing the same distribution for both RGs and GRGs, an exponential distribution reasonably describes the general features of both samples. The analysis also has some limitations. First of all, the two samples are selected at two different frequencies; in particular at 1.4 GHz the emission of cores is often more prominent and lobes 
 are usually more difficult to observe. Moreover, MB17 selected only sources with a flux density larger than 40 mJy at 1.4 GHz ($\sim$ 200 mJy at 150 MHz for $\alpha = 0.7$), while there are only 9 GRGs with a correspondingly larger flux density in 
 the BLDF sample. Furthermore, the BLDF sample is not complete in the range $0.7$ Mpc $<$ \textit{LLS} $< 1$ Mpc. These limitations might explain the difference of the inferred typical scales of the two samples.

\begin{table}
    \centering
    \begin{tabular}{p{1.1cm}p{3.2cm}p{3.2cm}}
    \hline
       & MB17 sample & BLDF-GRG sample \\ \hline
  Pareto & $c=0.02 \pm 0.01$,scale $=0.14 \pm 0.01$ & $c = 0.22 \pm 0.12$,scale$ = 0.52 \pm 0.03$ \\ 
         & $p=0.47$                            &  $p=0.94$ \\  
Exponential & scale $= 0.14 \pm 0.01$ & scale $=0.52 \pm 0.02$ \\ 
       & $p=0.52$                          & $p=0.70$ \\  
\hline
    \end{tabular}
    \caption{Results of the fit to the linear size distribution of the RGs provided from \citet{Miraghaei2017} and the BLDF-GRG sample. According to the p-value obtained by a KS-test, we cannot reject the hypothesis that exponential and Pareto distributions fit the \textit{LLS} distribution of both RGs and GRGs.}
    \label{tab:lls_test}
\end{table}


 \subsection{Redshift evolution and P-D diagram}
 \label{sec:z_evol}
 
\begin{figure}
        \centering
        \includegraphics[width= 0.5\textwidth]{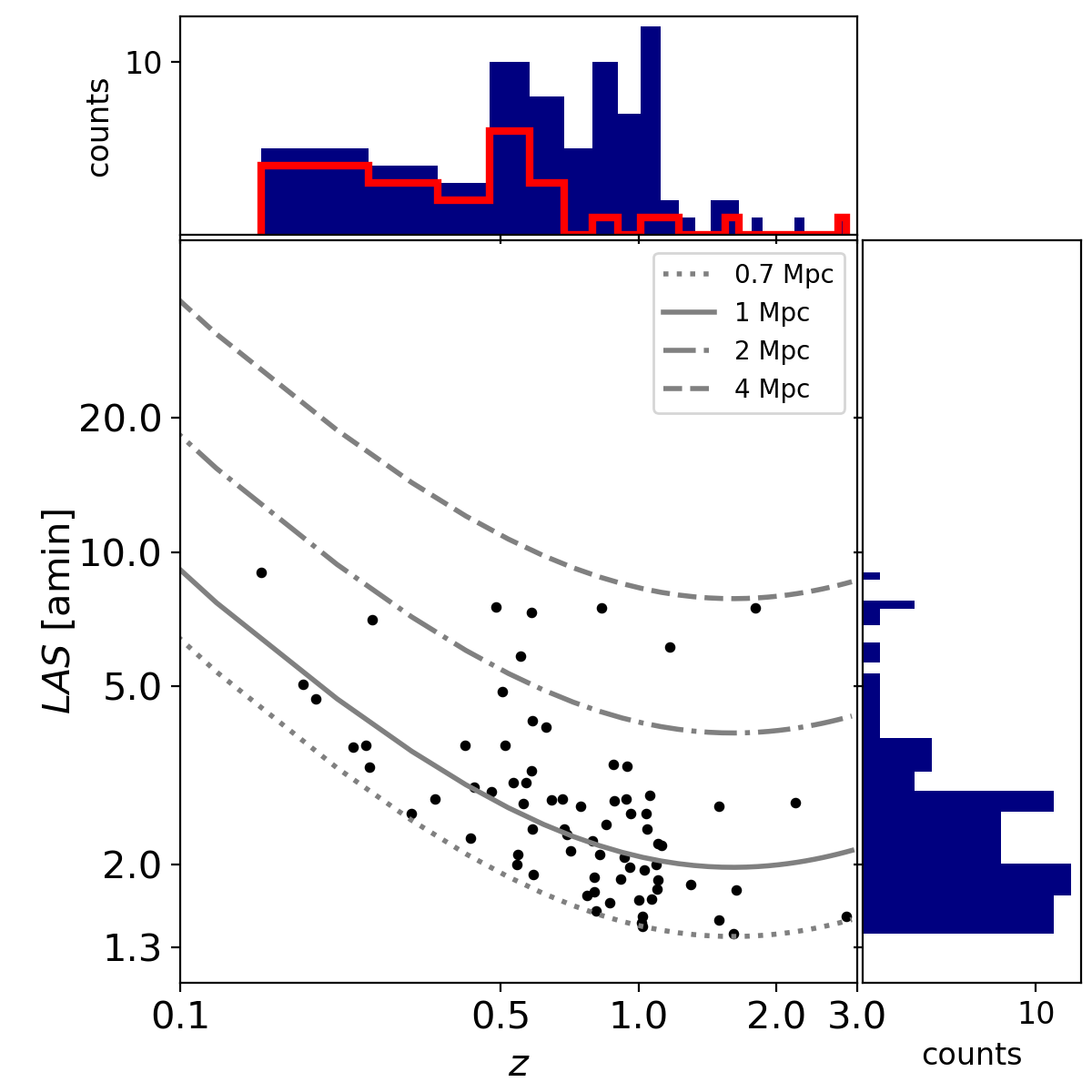}
        \caption{The \textit{LAS-z} diagram in the BLDF-GRG sample along with some reference lines of four "standard rulers", whose sizes are listed in the upper-right legend. Their distributions in redshift (upper panel) and \textit{LAS} (right panel) are also shown. The red histogram shows the spectroscopic redshifts.}
        \label{fig:LAS_vs_z}
\end{figure}

 We show the \textit{LAS-z} diagram for our BLDF-GRG sample in Fig.~\ref{fig:LAS_vs_z}, along with some reference lines that represent the angular size of the "standard rulers" of four different sizes as function of redshift. On the upper and right sides of Fig.~\ref{fig:LAS_vs_z} we show the distributions of redshifts and \textit{LAS}, respectively. The red histogram in the upper panel shows the distribution of the spectroscopic redshifts. It is worth emphasising that the minimum required angular size of RGs to be labeled as giant with \textit{LLS} $>$ 0.7 Mpc is 1.3$\rm '$ and 2.0$\rm '$ with \textit{LLS} $>$ 1 Mpc, and these minima occur near redshifts $\sim$1.7
 
\begin{figure}
        \centering
        \includegraphics[width= 0.5\textwidth]{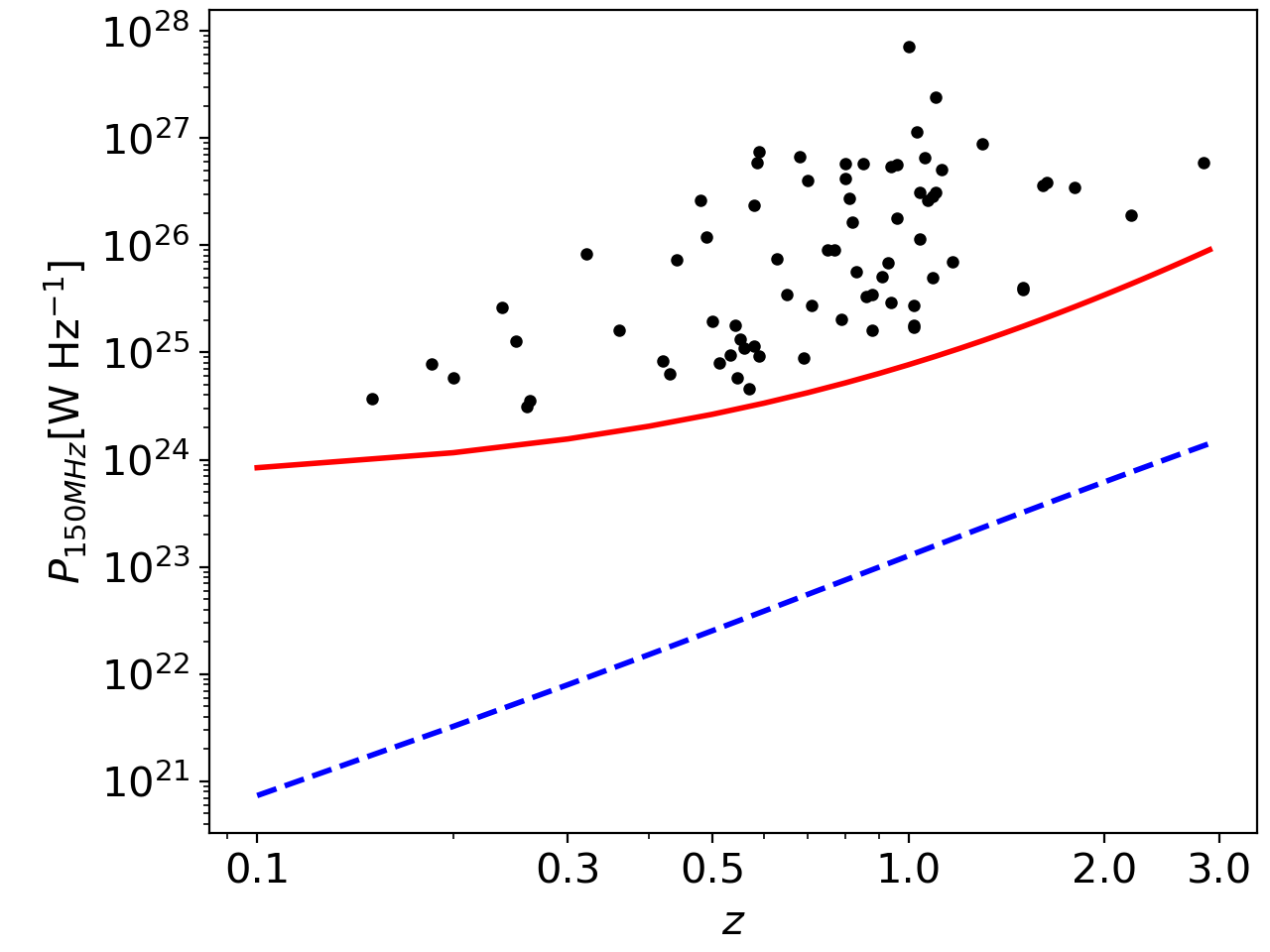}
        \caption{The relation between the radio power at 150 MHz, $P_{150}$, and the redshift, $z$, in  the BLDF-GRG sample. The solid red line shows the power of a GRG with a surface brightness equal to the noise level of the BLDF and integrated over its respective
        area of emission, at various redshifts. The dashed blue line traces the redshift evolution of the power of a point source with a flux density equal to 30 $\mu$Jy beam$^{-1}$.}
        \label{fig:p_vs_z}
\end{figure}

 In Fig.~\ref{fig:p_vs_z} we show the radio powers of BLDF-GRGs as a function of redshift. The solid red line shows the evolution of the power of a GRG with a surface brightness equal to 30 $\mu$Jy beam$^{-1}$ which is the noise level of the BLDF in the inner 3 $\rm deg^2$. We approximated the shape of GRGs by ellipses with a major and minor axis of 0.7 and 0.2 Mpc, respectively. The dashed blue line is the power of a point source with a brightness of 30 $\mu$Jy beam$^{-1}$ as function of redshift. These two lines correspond to the flux-limit of the Bo\"otes LOFAR survey and they provide a lower limit for the observed radio power. This result suggests that the clear trend we observe, with more powerful RGs at high redshifts, is likely due to a combination of the Malmquist bias and the surface brightness dimming proportional to $(1+z)^{-4}$ which is certainly important for extended or diffuse sources. Furthermore, inverse-Compton losses, due to interaction of the relativistic electrons with CMB photons will be larger than synchrotron radiative losses in the evolution of the lobes of giant radio sources \citep{Machalski2001, Konar2004}. Inverse-Compton losses are proportional to the initial energy of the CMB photon, leading to larger energy losses at high redshift due to the increasing CMB energy density which is proportional to $(1+z)^4$. Note also that in Fig.~\ref{fig:p_vs_z} there may be an artificial clustering at z$\approx$1 due to photometric redshift values in DESI DR9 rarely exceeding $z\approx$1.1.
 
\begin{figure}
        \centering
        \includegraphics[width= 0.5\textwidth]{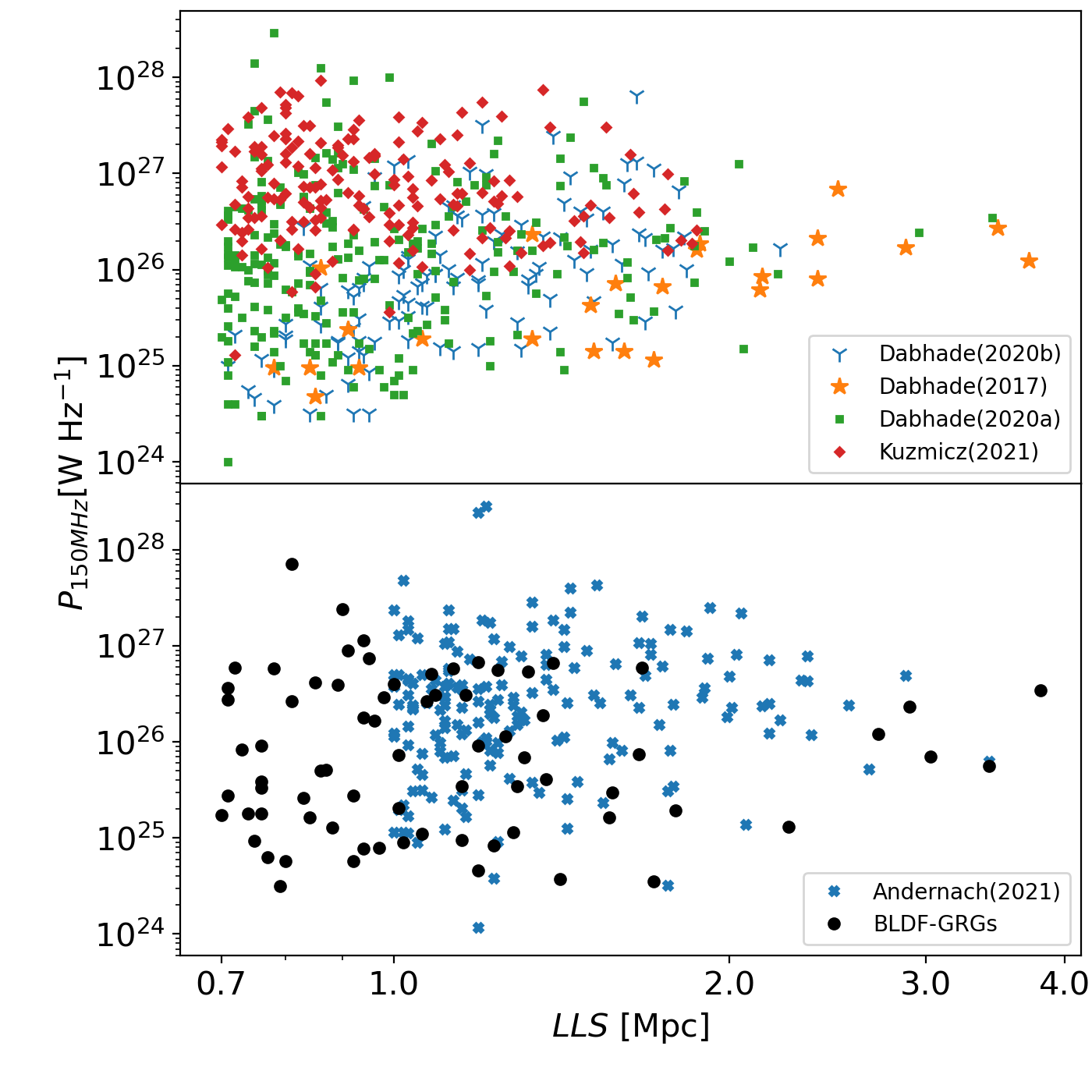}
        \caption{P-D diagram that shows the specific power at 1500 MHz versus the \textit{LLS} of the BLDF-GRGs. Other GRG samples are included for comparison, such as \citet{Dabhade2017}, \citet{Dabhade2020}, \citet{Kuzmicz2021}, \citet{SaganI} and \citet{Andernach2021}. The high sensitivity of the observations in this field enables us to detect several new GRGs at lower power ($\sim 10^{24} \rm W ~ Hz^{-1}$).}
        \label{fig:P_vs_D}
  \end{figure}

 The P-D diagram shows the relation between the radio power at a specific frequency and the linear size \citep{Baldwin1982}. With such a diagram it is possible to trace the evolution of RGs \citep{Ishwara-Chandra1999, Machalski2004}. The P-D diagram of our BLDF-GRG sample along with GRGs from recent samples \citep{Dabhade2017, Dabhade2020, SaganI, Kuzmicz2021, Andernach2021} is shown in Fig.~\ref{fig:P_vs_D}. We used a standard spectral index, $\alpha = 0.7$, to compute the power at 150 MHz for those sources with a flux estimated at other frequencies. As found before \citep[e.g.,][]{Ishwara-Chandra1999, Kuzmicz2012, Dabhade2020, Kuzmicz2021}, the larger RGs are less powerful. A deficit of GRGs is clearly visible in the upper-right corner, where powerful and large RGs should reside, suggesting that the luminosity of RGs decreases as they evolve to giant radio sources which is likely the latest stage of the evolution according to models \citep[e.g.,][]{Kaiser1997}. The electron energy losses due to adiabatic expansion and radiation over the lifetime of the lobes could cause such a deficit. Fig.~\ref{fig:P_vs_D} also shows a lack of GRGs with large linear sizes and small radio powers. This result might suggest that RGs have to be powerful enough in order to reach the largest sizes. However, the most extended RGs have a rather low surface brightness, thus their detection becomes more difficult with increasing size. As a matter of fact, we have been starting to observe large ($>$ 1 Mpc) and faint ($\sim 10^{24} \rm W ~ Hz^{-1}$) GRGs only with the most recent observations with LOFAR and ASKAP \citep{Andernach2021}.
 
 \subsection{Spectral indices}
 
Radiative ageing affects the radio spectra since more energetic electrons suffer higher energy losses and, consequently, have shorter lifetimes. This induces a steepening of the spectra at high frequencies ($\gtrsim$ GHz) and the spectral index can give an estimate of the age of a radio source. 
In order to compute the integrated spectral index (i.e., estimated from the total flux of the source) we performed the following steps: First, we convolved the higher-resolution images (LoLLS and LBDF) to the (lower) resolution of 45$\rm "$ of NVSS and regridded the images to a size of 0.2$^\circ$ centered on the sky coordinates of the source. This step is necessary because the measured flux can change with the resolution of the image. We performed sigma-clipping at a level of $3\sigma_{\rm rms}$, where $\sigma_{\rm rms}$ is the rms noise of the individual convolved image, and integrated the flux of the sources. If the source is undetected in the NVSS, we considered an upper limit for the flux of such source equal to $3\sigma_{\rm rms}$. For sources that are detected neither in NVSS nor in LoLSS we did not estimate the spectral index. We also excluded those GRGs that were blended with other sources after convolution. Once the fluxes at different frequencies are calculated, we performed a linear regression to calculate the spectral indices. In order to compute a more robust error, we carried out a simple bootstrapping to determine the errors in the fit parameters. This routine generates Gaussian-distributed random fluxes whose mean is equal to the initially estimated flux and a standard deviation that is determined from the error of this flux. Then, a fit is performed for each dataset and the final variance of the multiple fits is used as the error of the spectral index. 
 
\begin{figure}
        \centering
        \includegraphics[width= 0.5\textwidth]{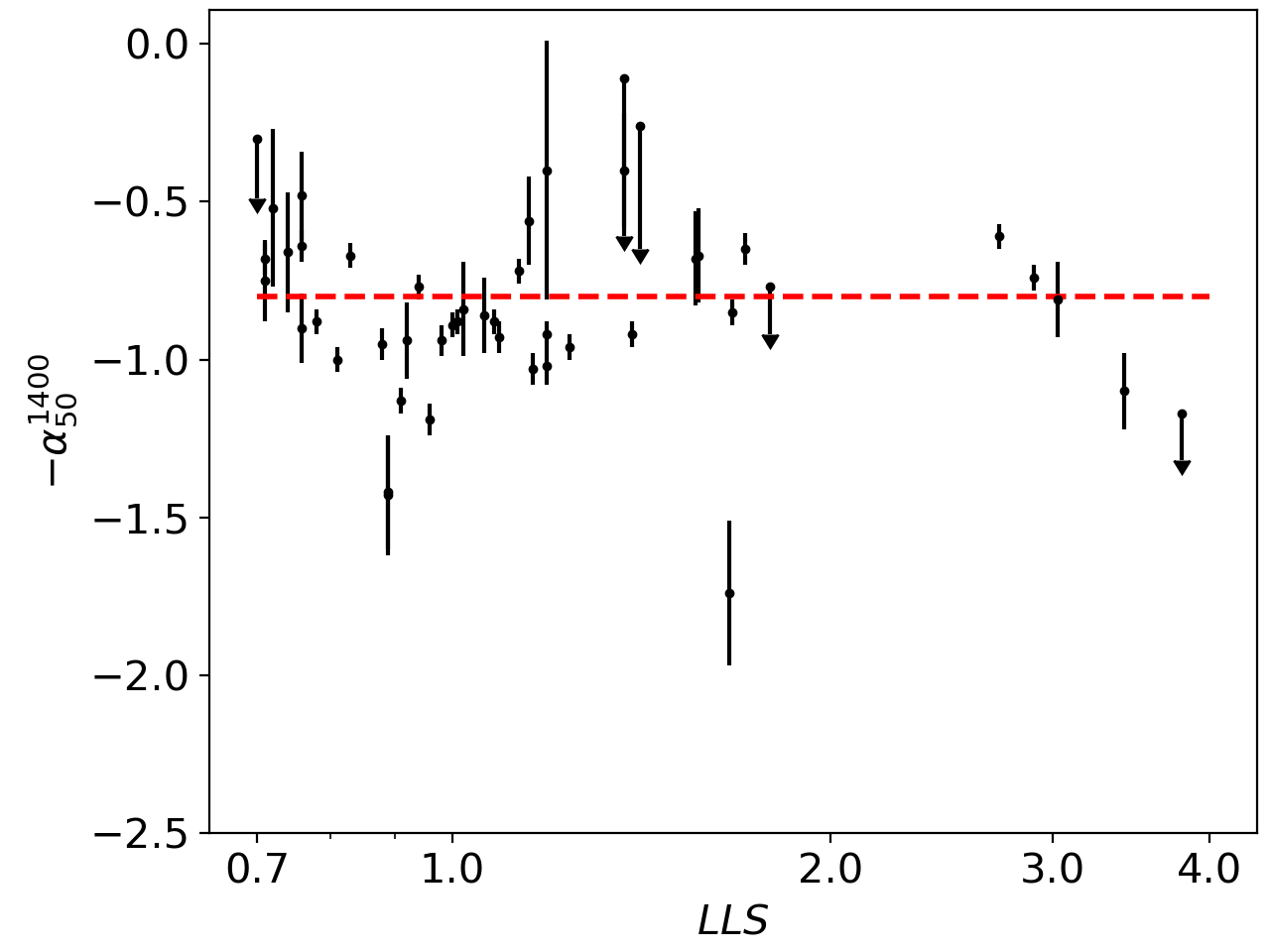}
        \caption{Spectral index in the range 50-1400 MHz for the BLDF-GRG sample against the linear size. The typical slope of radio spectra in such RGs is about -0.8, which is denoted by the red line. GRGs do not show a substantial steepening at the largest linear extent, mainly because we detect the regions with recently injected or re-accelerated particles (e.g., hotspots).}
        \label{fig:spidx_vs_LLS}
  \end{figure}

\begin{figure}
        \centering
        \includegraphics[width= 0.5\textwidth]{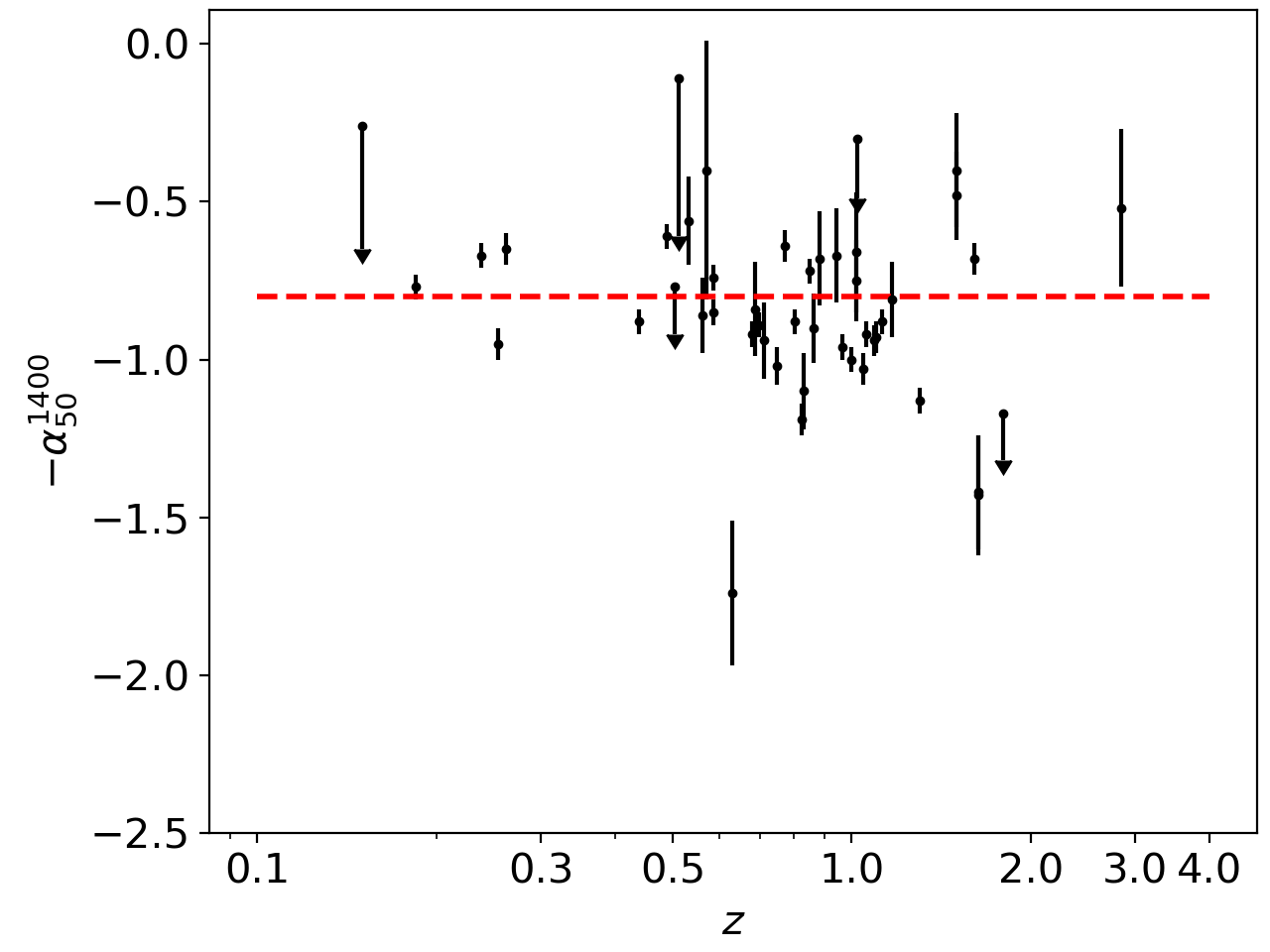}
        \caption{Spectral index in the range 50-1400 MHz for the BLDF-GRG sample against the redshift of the host galaxies. The spectral index does not show a clear trend with redshift.}
        \label{fig:spidx_vs_z}
  \end{figure}

\begin{figure}
        \centering
        \includegraphics[width= 0.5\textwidth]{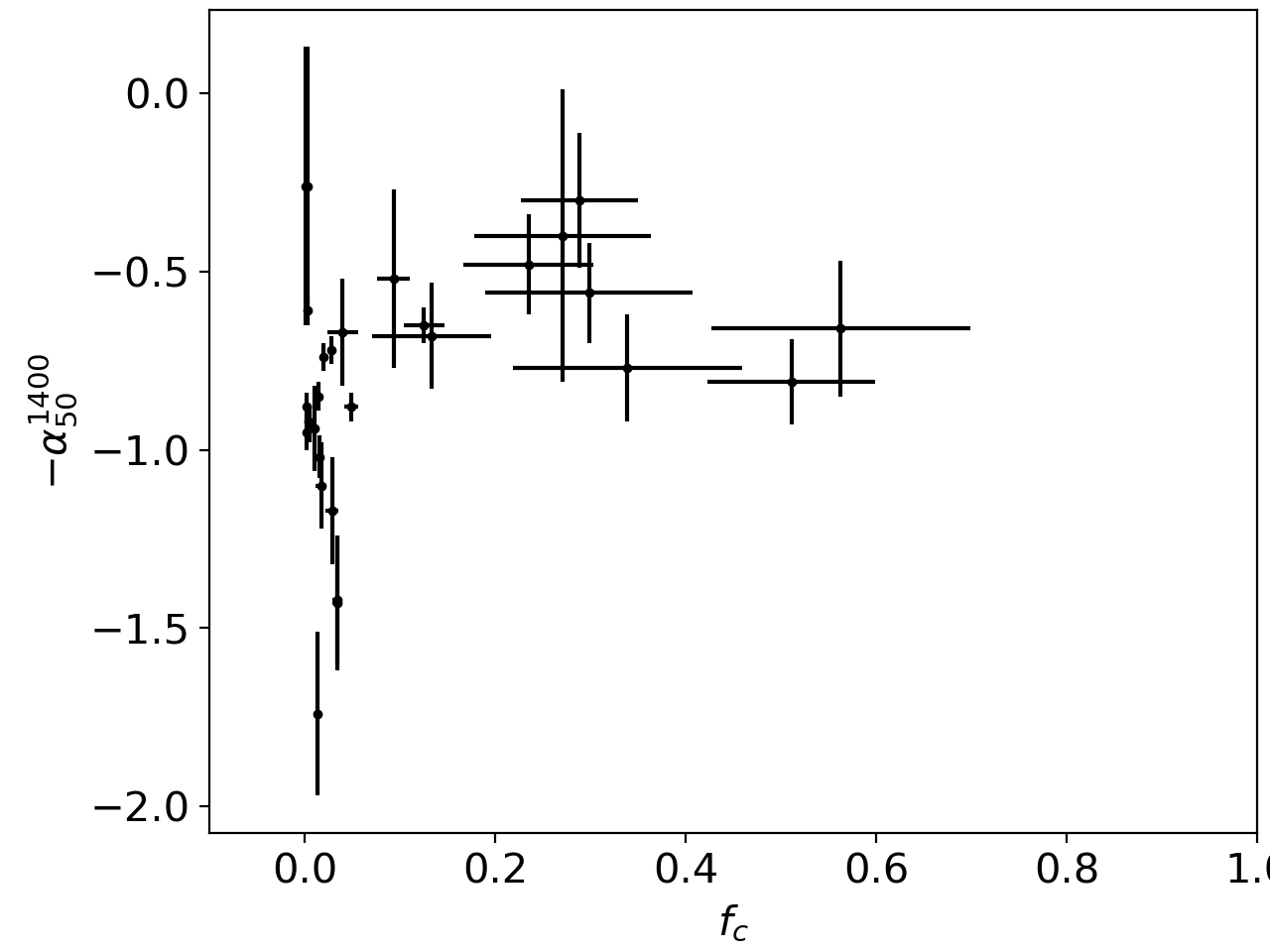}
        \caption{Spectral index in the range 50-1400 MHz for the BLDF-GRG sample against the core fraction $f_c$. Due to the increasing contribution of the core to the total emission, the radio spectra flatten in the range $0 < f_c < 0.2$, while the spectral index is constant for larger values of the core fraction.}
        \label{fig:spidx_vs_fcore}
  \end{figure}

 In Fig.~\ref{fig:spidx_vs_LLS} we show the relation between the spectral index and the linear size in our BLDF-GRG sample. 
 We find spectral indices around 0.8 and no substantial spectral steepening with  increasing \textit{LLS}, mainly because the integrated flux is dominated by the regions with recently injected or re-accelerated particles (the core and hotspots). Moreover, Fig.~\ref{fig:spidx_vs_z} does not suggest a relation between the spectral index and the redshift. On the other hand, \citet{SaganI} recently found a weak correlation between the spectral index and linear size and the redshift in a larger sample of GRGs, in agreement with \citet{Blundell1999} and \citet{Jamrozy2008}. Large and complete samples are needed to establish whether GRGs exhibit a correlation between these properties, as well as high-resolution spectral index maps to trace a possible steepening of the spectra along the lobes.
 We also discuss the contribution of the core to the total flux at 150 MHz, that is the core fraction, $f_c$, defined as the ratio between the flux of the core and the total flux. The emission of the core of the RGs is dominated by the recent ejecta launched by the central black hole and it usually has a flatter ($\alpha < 0.5$) or inverted spectral index ($\alpha < 0$) \citep{Konar2004, Konar2008}. This would in turn flatten the integrated spectral index calculated from the total flux of the radio source. Thus, we calculated the flux of the core of GRGs in the BLDF image. In case of absence of core emission or indistinguishable core, we did not calculate the flux. Fig.~\ref{fig:spidx_vs_fcore} shows the integrated spectral index against the core fraction. For better visualisation, all spectral indices are treated as absolute measurements with a relative error, irrespective of whether they are upper limits or not. An expected trend of the spectra becoming flatter for larger values of $f_c$ can be seen and for $f_c\approx 0.2$ the spectral index is about $0.5$. This indicates a flattening of the integrated spectral index induced by the relatively large contribution of the emission of the core of some GRGs. Such a core fraction suggests that most GRGs are not entirely passive and might be in a phase of restarted activity. 
 
  \subsection{HERG and LERG dichotomy}
  
  Based on their optical spectra, AGN can be classified as High-Excitation Radio Galaxies (HERG) and Low-Excitation Radio Galaxies (LERG). The former have an accretion rate onto the black hole between one and ten per cent of the Eddington ratio. They are hosted by bluer, star-forming galaxies and lower-mass black holes. In contrast, LERGs are likely hosted by high-mass galaxies with a central black hole that experiences accretion below one percent of the Eddington limit \citep{Best2012}. 
  
\begin{figure}
        \centering
        \includegraphics[width= 0.5\textwidth]{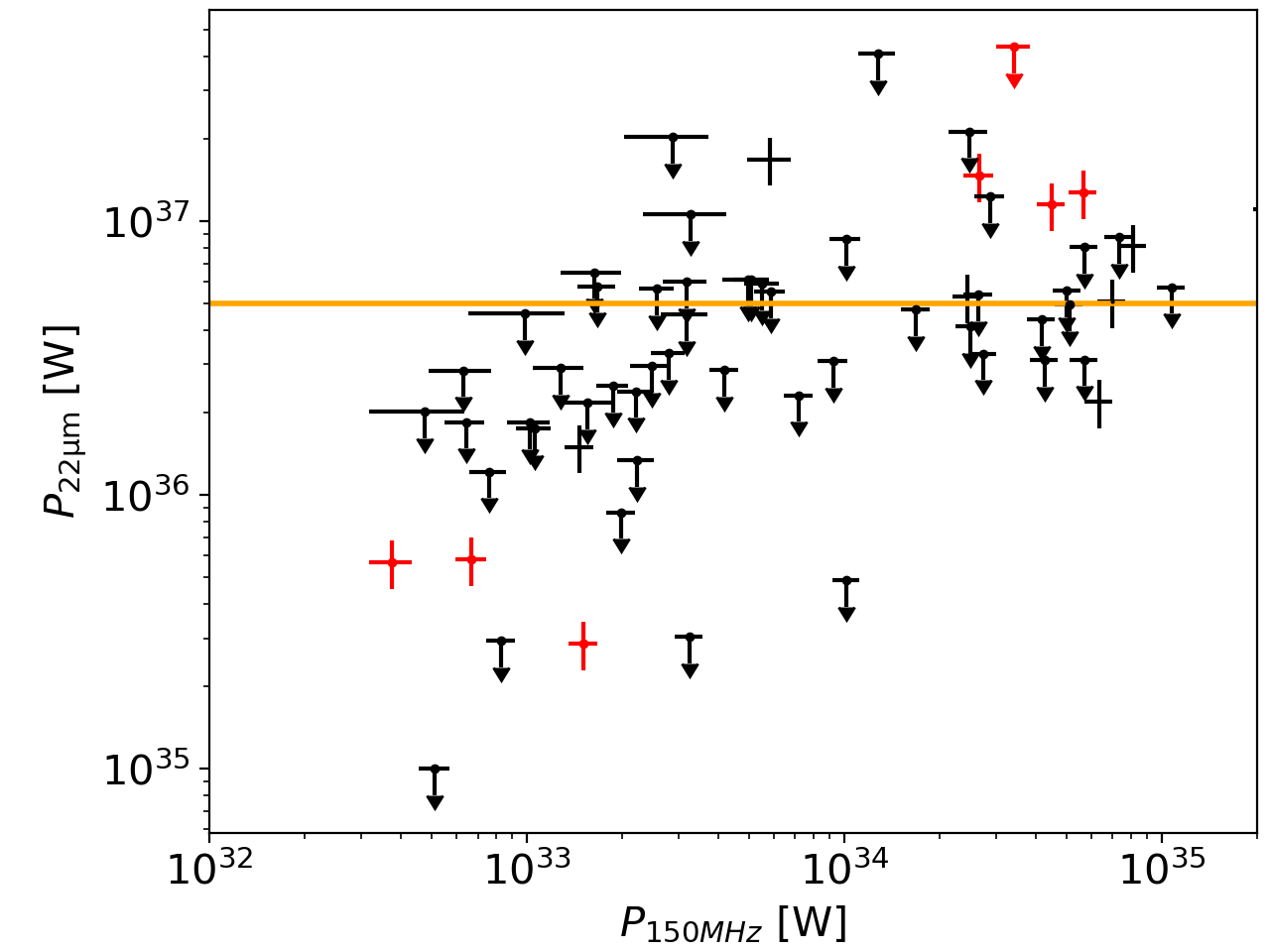}
        \caption{Infrared ($22 \mu$m, W4) - radio (150 MHz) power plot for BLDF-GRG sample. The arrows are upper limits with a signal-to-noise ratio lower than 2 in the W4 band. The orange line is an empirical boundary between HERGs and LERGs \citep{Gurkan2014}. The plot indicates a predominance of LERGs in our sample. The red points are either spectroscopically identified quasars or candidate quasars. }
        \label{fig:PIR-PR}
  \end{figure}

\begin{figure}
        \centering
        \includegraphics[width= 0.5\textwidth]{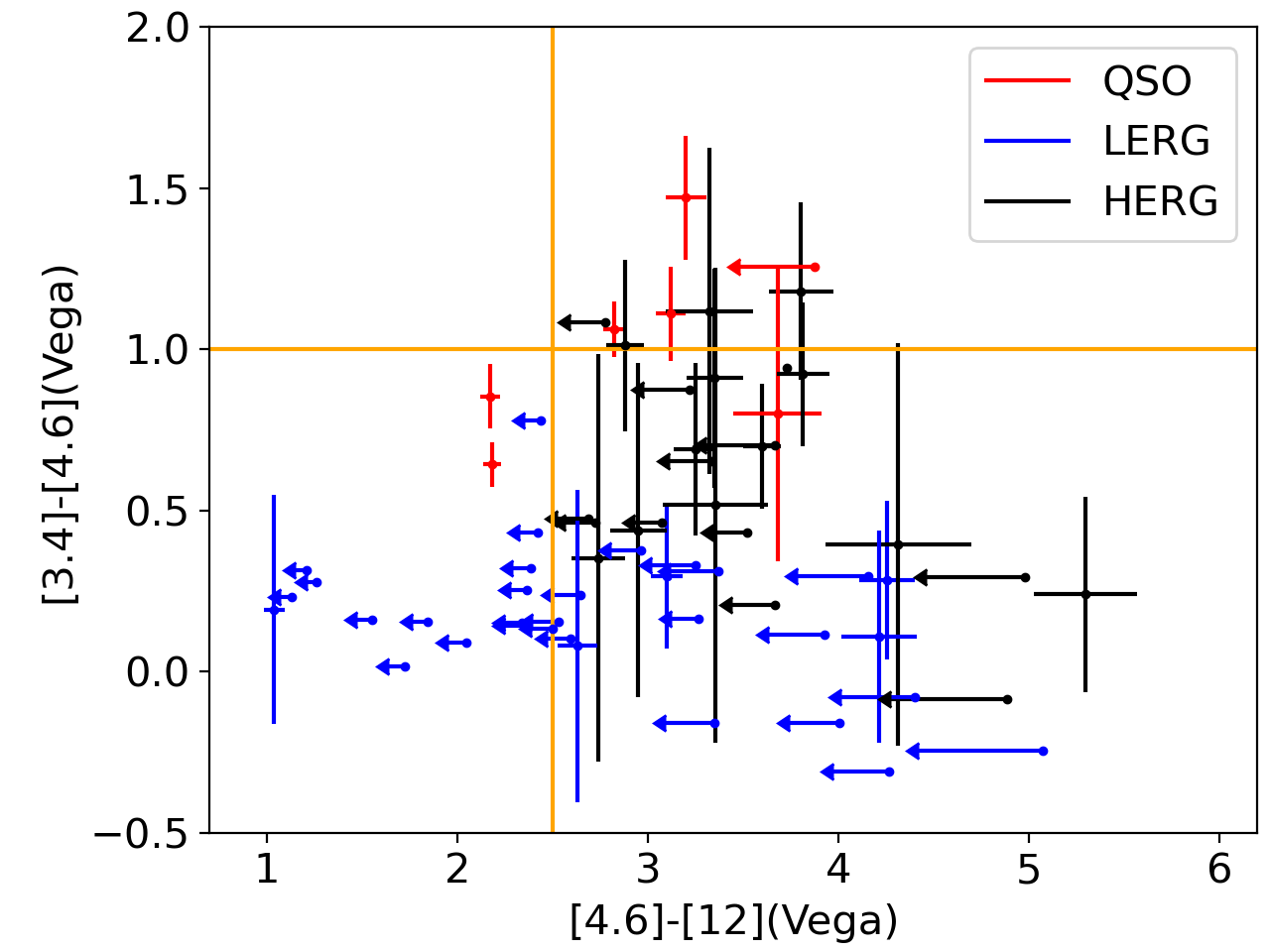}
        \caption{WISE colour-colour diagram in the BLDF-GRG sample. Red dots mark either spectroscopically identified quasars or candidate quasars. Almost every GRG host lies in the HERG region of the plot. However, half of them are upper limits (arrows), which means that the source has a signal-to-noise ratio smaller than 2 in the W3 band. The classification as HERGs or LERGs is based on the infrared power relation criterion (Fig.~\ref{fig:PIR-PR}).}
        \label{fig:WISE_colour-colour_plot}
  \end{figure}

  The host galaxies in our BLDF-GRG sample are usually too faint to allow a SED fitting and provide a classification based on the optical spectra. Nevertheless,  mid-IR data can be used to classify AGN \citep{Assef2010, Jarrett2011, Stern2012, Mateos2012, Assef2013}. \citet{Gurkan2014} have shown that HERGs and LERGs have different mid-IR luminosities. The former are mostly luminous sources in the infrared band and an empirical cutoff can be drawn in the $22 \mu$m - 151 MHz luminosity plot (e.g., Fig.~\ref{fig:PIR-PR}), with LERGs showing a mid-IR luminosity below $4-5 \times 10^{43} \rm ~ erg\, s^{-1}$ ($4-5 \times 10^{36} \rm ~ W$). Moreover, while LERGs lie in the bottom-left region of the WISE colour-colour plot (which is shown in Fig.~\ref{fig:WISE_colour-colour_plot}), HERGs are mostly located on the right side. The $y$-axis is the difference between the magnitude in the W1 and W2 band, while the $x$-axis shows the colour obtained from W2 and W3. Fig.~\ref{fig:PIR-PR} shows the relation between the radio power at 150 MHz and the infrared power at $22 \mu$ m for the BLDF GRGs. It is worth noting that most of our sources have values or upper limits for the infrared power below the orange horizontal line, indicating that in our sample the optical host galaxies are predominantly LERGs. This is also confirmed by the WISE colour-colour plots (Fig.~\ref{fig:WISE_colour-colour_plot}; W2-W3 vs W1-W2). In this figure, we coloured the host galaxies according to the classification into HERGs (black) and LERGs(blue) suggested from Fig.~\ref{fig:PIR-PR}. Even though only a few sources lie in the region delimited by W1 - W2 $<$ 1 and W2 - W3 $<$ 2.5 (namely the LERGs region), almost all of them are upper limits (arrows), in agreement with Fig.~\ref{fig:PIR-PR}, and may shift into the LERG region with deeper mid-infrared observations. A source has an upper limit in the infrared power if the signal-to-noise ratio is lower than 2, either in the W3 or W4 band. The red points in both plots are either spectroscopically identified quasars or quasar candidates in our sample and lie in the upper region (W1 -W2 $\gtrsim 1$ and W2 - W3 $\gtrsim 2 $).
  Our sample is drawn from the faintest sources, which poses a limitation to our analysis since most of the GRG hosts are not detected in the W4 band. 
  
   Best et al. (subm.) and \citet{Mingo2022} provide an AGN classification into HERG and LERG performed by SED fitting in the optical band. Therefore, we validated our results by crossmatching our BLDF-GRG sample with these catalogues. We found 35 matching GRGs, of which only ten are radiatively efficient AGN. Recently, \citet{Mingo2022} reported that the majority of FRII galaxies experience low accretion rates, especially for the low-power ($L_{150} < 10^{26} \rm~ W ~ Hz^{-1}$) AGN population. Here, we confirm this result for GRGs as most of them seem to be in a radiative inefficent mode, regardless of the radio power. Based on the excitation state, it does not appear that GRGs are undergoing massive, radiatively efficient accretion at the present time. However, it is possible that the central engine has gone through a series of recurrent AGN events that have allowed the GRGs to grow to its extreme size.
   
\begin{figure}
        \centering
        \includegraphics[width= 0.5\textwidth]{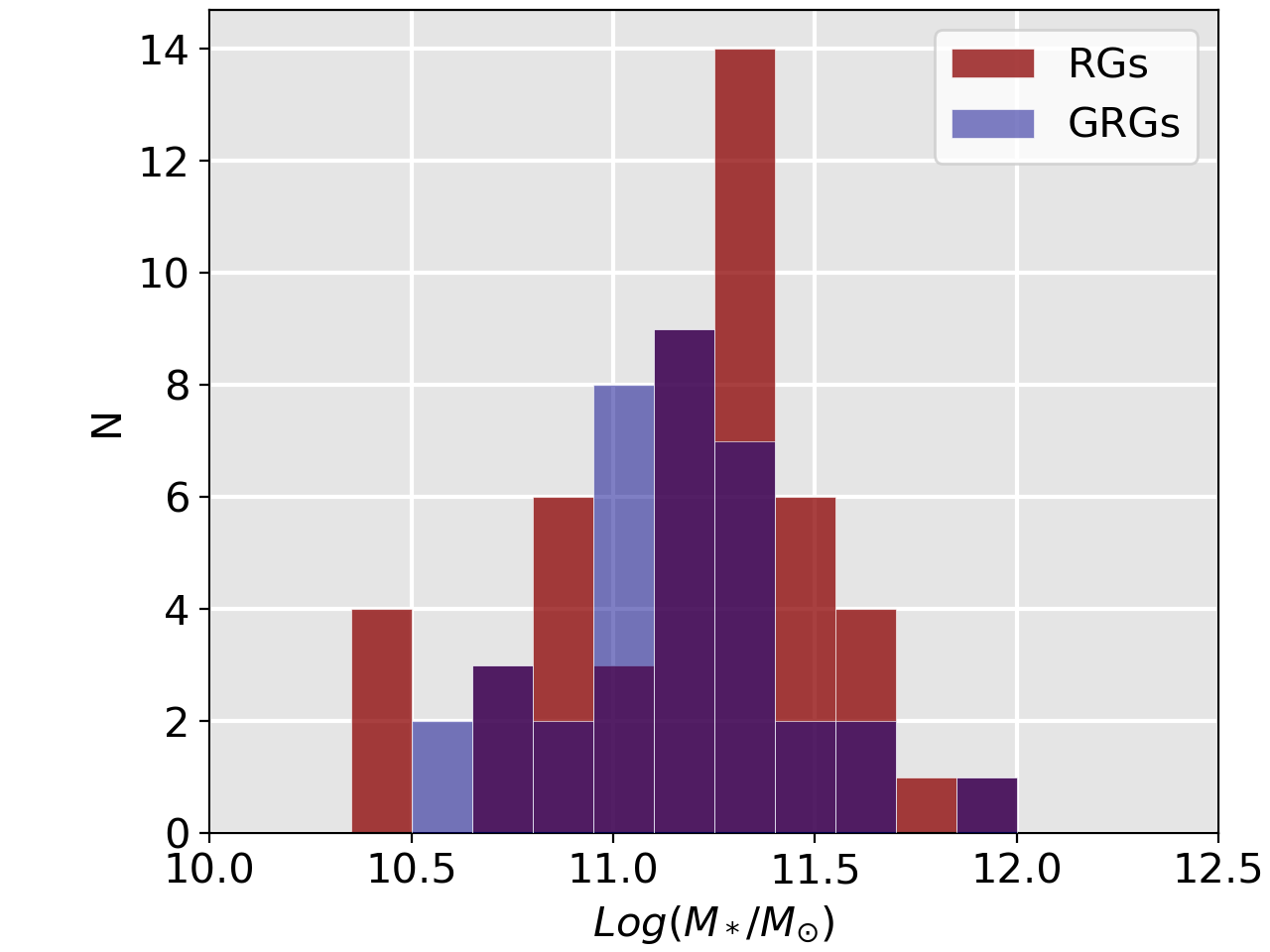}\quad \includegraphics[width= 0.5\textwidth]{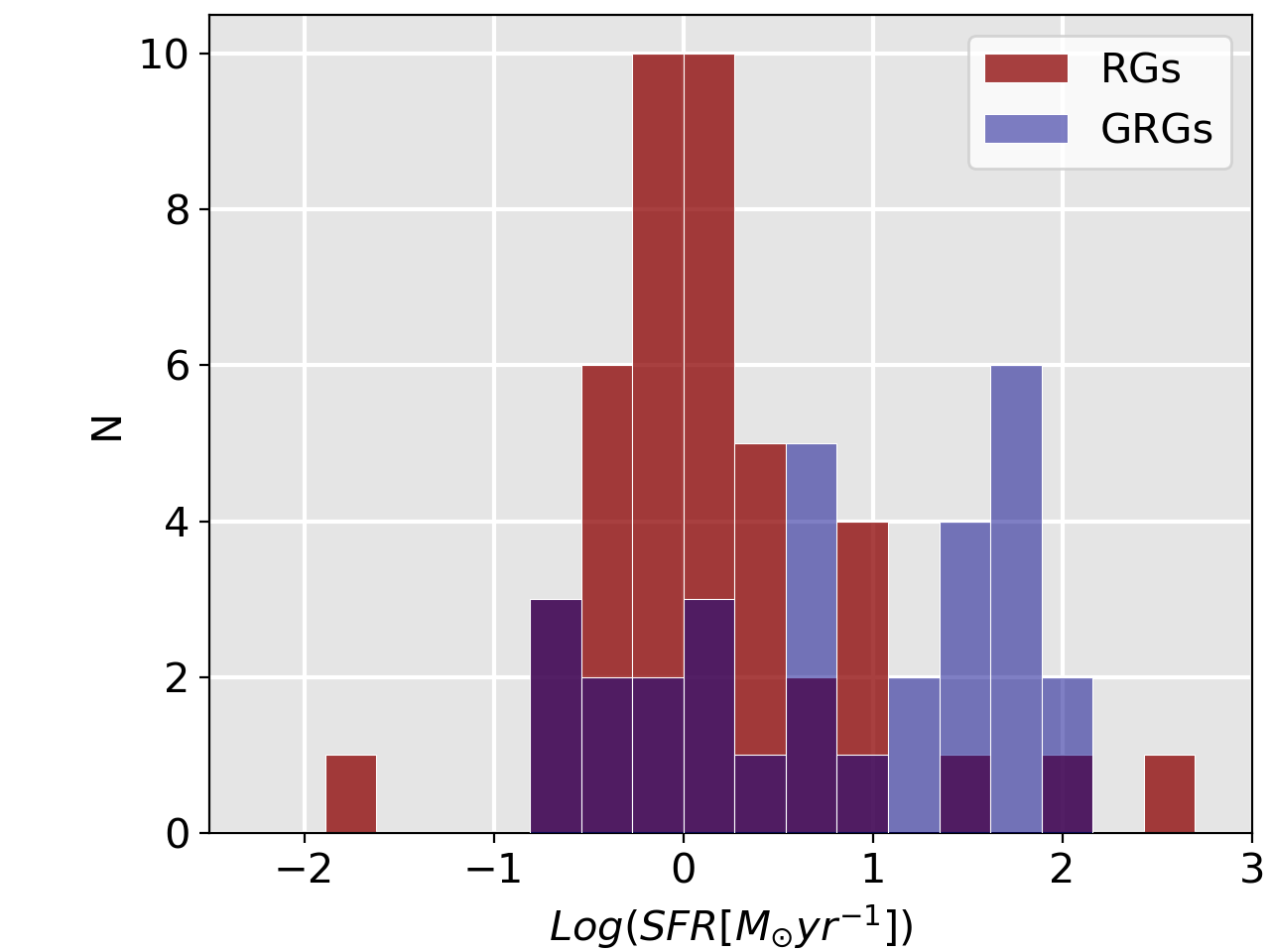}
        \caption{Upper panel: Distribution of the stellar mass in GRG (blue) and RG (red) hosts. The former are usually very massive galaxies ($> 10^{10.5} M_{\odot}$), while the latter that span the entire range from $10^{10}$ to $10^{12} M_{\odot}$.
        Lower panel: SFR distribution for GRGs (blue) and RGs (red). The distribution of GRGs is uniform, indicating star formation activity in at least 10 out of 23 GRG hosts.}
        \label{fig:stellar_mass_distribution}
  \end{figure}

   During the inspection of the BLDF, we found 152 extended RGs which have a  linear size in the range 10-700 kpc. We selected the RGs by angular size, trying to be reasonably complete for \textit{LAS} $\gtrsim$ 1.3$\rm '$. For this reason, only 3 of them have a linear extent smaller than 100 kpc.
   We compared the stellar mass and star formation rate (SFR) of GRGs and smaller radio galaxy (RG) hosts in Fig.~\ref{fig:stellar_mass_distribution}. The total sample contains 243 sources (GRGs+RGs), 87 (35 GRGs and 52 RGs) of which have either SFRs or stellar mass values from either Best et al. (subm.) or \citet{Mingo2022}. In our sample, GRG hosts have high stellar masses ($> 10^{10.5} M_{\odot}$) and the median decimal logarithm of the stellar mass for RGs (11.24) and GRGs (11.13) is similar. Nevertheless, the distribution of GRG hosts shows an excess of stellar stellar masses lower than $\rm M_* = 10^{11.5} \rm M_{\odot}$ with respect to RG hosts. The result is in agreement with previous studies which claimed that GRGs are usually hosted by luminous elliptical galaxies dominated by the emission of evolved giant stars \citep{Lara2001, Machalski2001, Clarke2017, Dabhade2017, Seymour2020}. These are effectively dead systems in which most of the star formation and black hole growth have already come to an end, suggesting that GRGs are the last stage of the evolution of RGs. However, very rare, massive spiral galaxies that host relativistic jets and lobes that extend to Mpc scale have been found as well \citep[e.g,][and reference therein]{Mao2015}.  Moreover, \citet{Kuzmicz2019} performed a detailed analysis of the star formation history in a sample of GRGs finding an 'intermediate' stellar population with an age between $9 \cdot 10^8 - 7.5 \cdot 10^9$ yr,  besides the common old stellar population residing in the host galaxies ($t > 10^{10}$ yr). Similar results have been found in the GRG ESO 422–G028 by \citet{Zovaro2022}. This is indicative of star formation activity at least in some GRG hosts, as shown in the lower panel of Fig.~\ref{fig:stellar_mass_distribution}. In fact, we see an excess of GRGs at SFR values of 10-100 $ \rm M_{\odot} ~ yr^{-1}$ with respect to RGs.

  \subsection{GRG environment}
  
  Several studies have addressed the role of the environment in the evolution of RGs in the past. It is known that only a small percentage of GRGs reside in galaxy clusters or groups \citep[e.g.,][]{Dabhade2020, Andernach2021} and it is commonly believed that the intracluster medium can frustrate the expansion of jets and lobes in the local environment \citep{Subrahmanyan2008, Safouris2009}. 
  
\begin{figure}
        \centering
        \includegraphics[width= 0.5\textwidth]{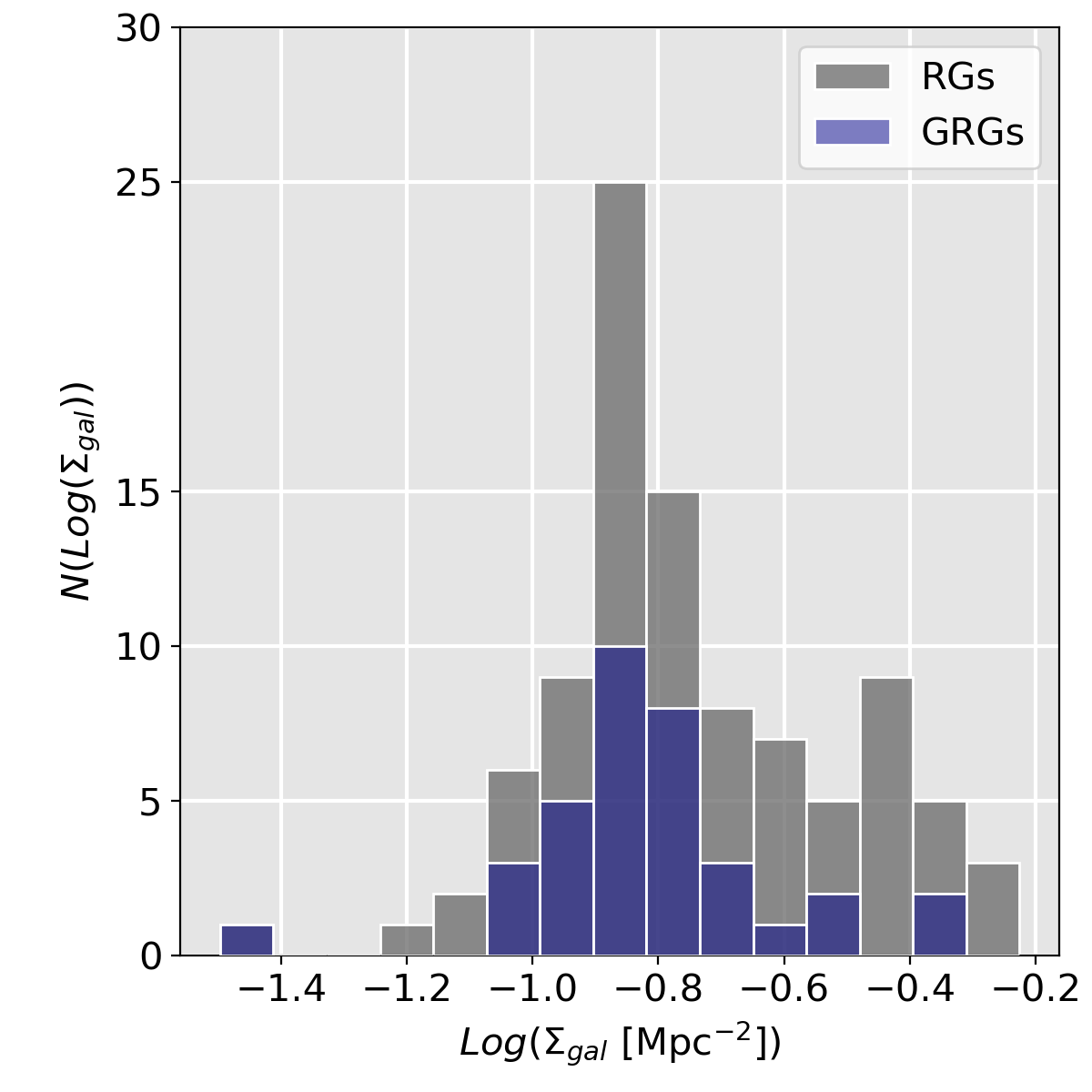}
        \caption{Distribution of the surface number density of galaxies within 10 Mpc of the GRGs (blue) and RGs (grey). According to the KS test, we cannot rule out the hypothesis that the samples are drawn from the same distribution at a level of significance of 90\%. Nevertheless, the figure shows that GRGs have a tendency to reside in underdense environment.}
        \label{fig:sigmagal_distribution}
  \end{figure}

  We tested the conjecture that GRGs are preferentially located in underdense environments using photometric redshifts from the DESI Legacy Imaging Survey \citep{Legacy2019}, assuming that the galaxies trace the distribution of the intergalactic medium. The DESI DR9 photometric redshift catalogue includes observations from the Beijing-Arizona Sky Survey \citep[BASS,][]{BASS2017}, DECam Legacy Survey (DECaLS) and Mayall z-band Legacy Survey (MzLS) \citep{Legacy2019}.
  In order to look for differences in the environments of GRGs and ordinary RGs, we performed this analysis for our BLDF-GRG sample, as well as for the sample of smaller RGs we found in the BLDF. Due to the depth limit of the DESI survey, we restricted our analysis to $z<0.7$ yielding 35 GRGs (23 of which have linear sizes larger than 1 Mpc) and 96 RGs. To examine the source environments, we first have to create a volume-limited sample, by cutting out those galaxies with an absolute magnitude (luminosity) in the $r$ band brighter than the absolute magnitude (luminosity) of a galaxy with an apparent magnitude equal to the flux limit, in the $r$ band, located at the maximum redshift considered ($z=0.7$). Such a cut limits our analysis since it is based on the most luminous galaxies which are the rarest as well. As a consequence, we might miss galaxies that are members of galaxy clusters and understimate the overdensities. We then performed a similar analysis dividing the sample into three redshift bins, $0 \le z < 0.3$, $0.3 \le z < 0.5$, $0.5 \le z < 0.7$. This method enables us to utilise a larger number of galaxies at low redshifts and increase the robustness of the measurement of the overdensities. The results of the two analyses are similar and, due to the poor statistics of the second method, we decide to report the findings of the first method described. 
  We counted the number of galaxies within a sphere with a radius equal to a comoving distance of 10 Mpc centred on the coordinates of the (G)RG hosts. We approximated the redshift of the source with the redshift value reported in Tab.~\ref{tab:list}, without taking into account the error. We show the distribution of GRGs (blue) and RGs (grey) with surface number density of galaxies, $\Sigma_{\rm gal}$, within 10 Mpc in Fig.~\ref{fig:sigmagal_distribution}. When testing for differences in the galactic environments around GRGs and RGs, the KS test yields a p-value of 0.17. This result shows that there is no evidence that GRGs reside in environments different from other RGs, in agreement with \citet{Komberg2009} and \citet{Lan2021}.
  
\begin{figure}
        \centering
        \includegraphics[width= 0.5\textwidth]{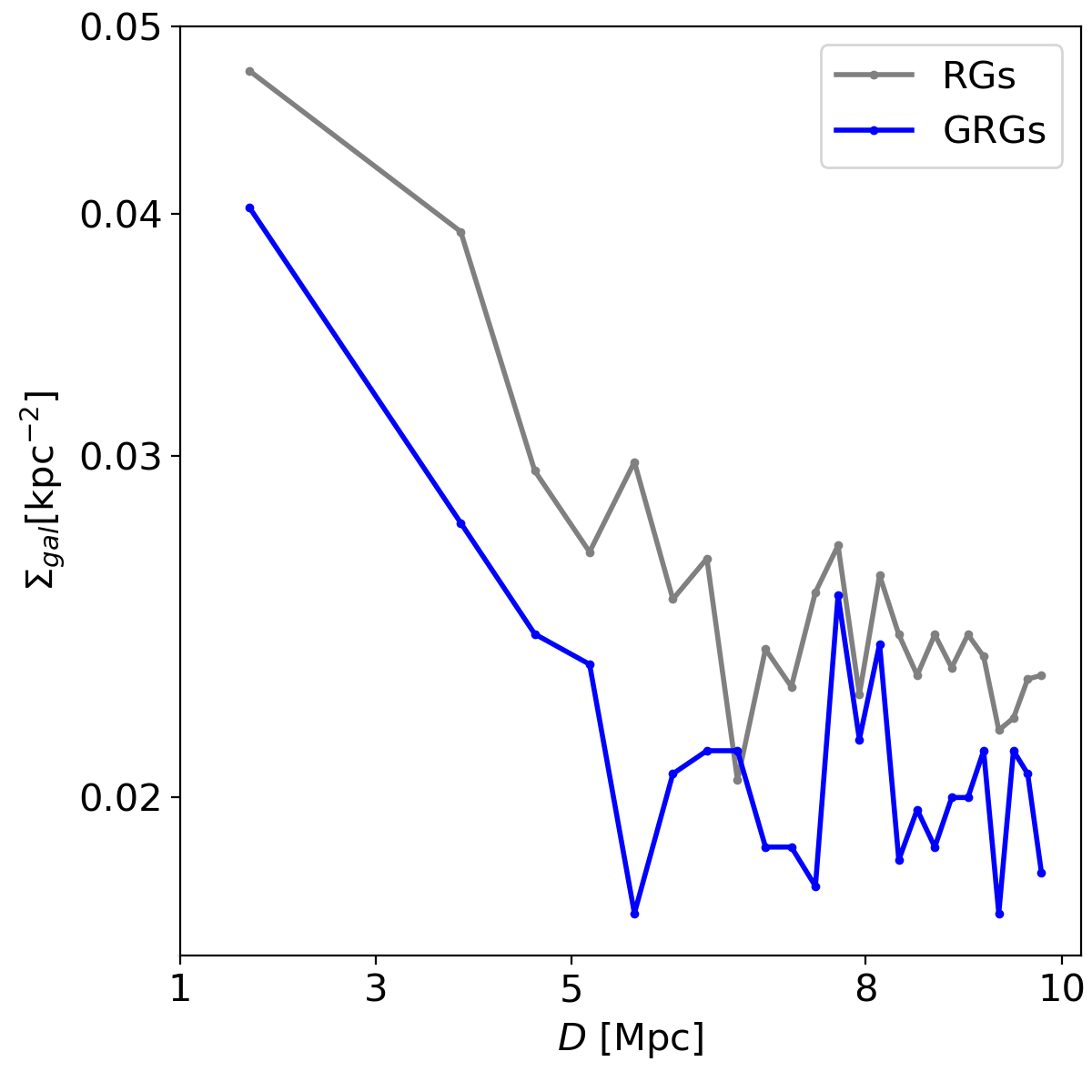}
        \caption{Radial distribution of the surface number density of galaxies within 10 Mpc of the GRGs (blue) and RGs (grey). The KS test suggests that the two populations inhabit similar environments.}
        \label{fig:sigmagal_radial}
  \end{figure}

  Dividing the sphere of radius 10 Mpc into 15 shells with equal volumes, we compute the radial profile of the surface number density of the galaxies (Fig.~\ref{fig:sigmagal_radial}). GRGs have a tendency to reside in sparser environment with respect to ordinary RGs as the latter have a larger number density of galaxies at almost every radius. However, the p-value of 0.83 of a KS test comparing GRGs with ordinary RGs suggests that this result is not statistically significant. 
 
\begin{figure}
        \centering
        \includegraphics[width= 0.5\textwidth]{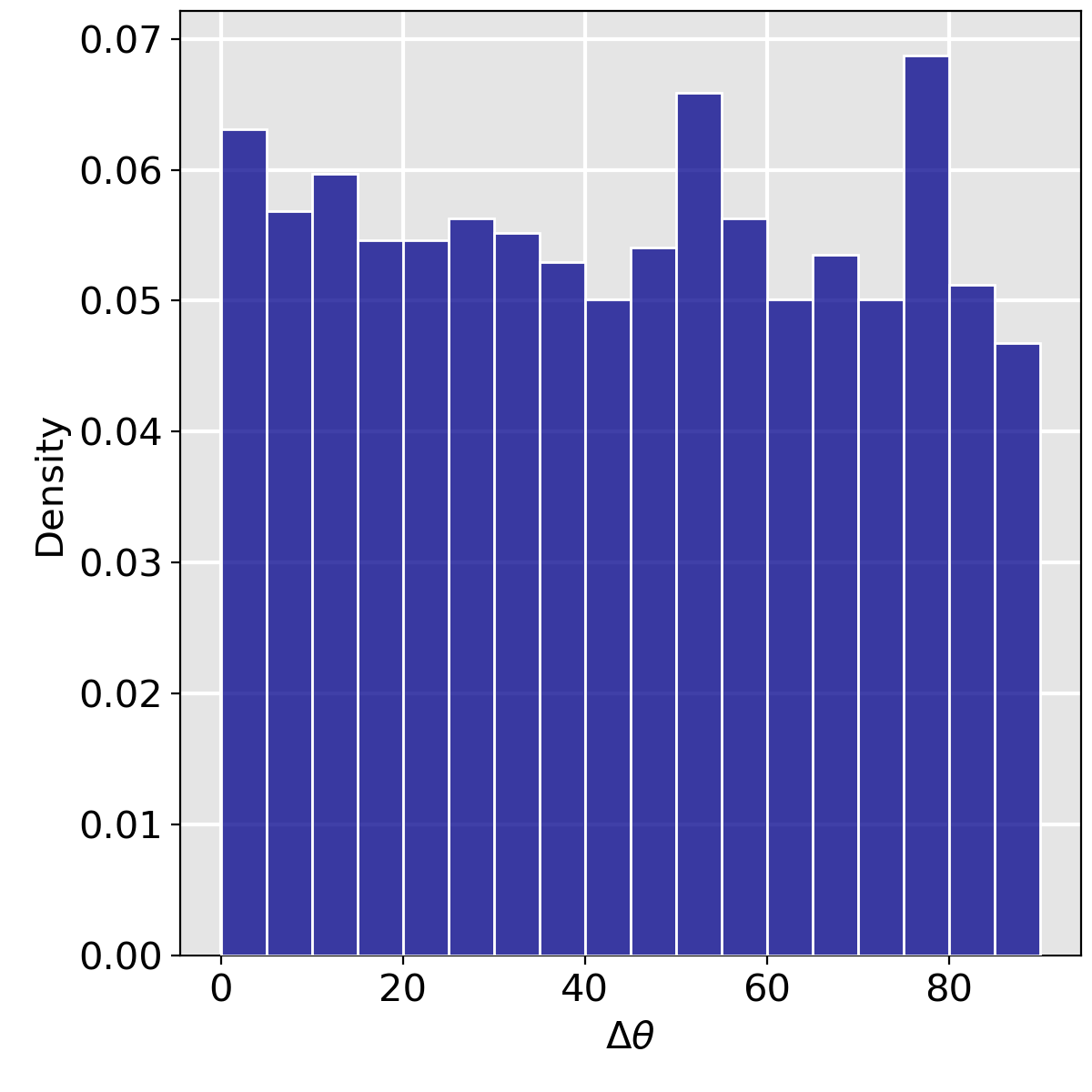}
        \caption{Distribution of the orientation of galaxies surrounding GRGs with respect to the direction of the expansion of jets. The distribution is consistent with being uniform.}
        \label{fig:deltapa_distribution}
  \end{figure}

   Our results are in agreement with \citet{Alcyneous2022}, who found the largest GRG known to date (\textit{LLS}=5 Mpc, $z=0.25$) and it has only 5 galaxies within 10 Mpc, leading to a surface number density  of 0.05 Mpc$^{-2}$, and does not have a single galaxy within 5 Mpc. The value of the surface number density and the number of the galaxies are similar to those reported in \citet{Lan2021}, albeit a direct comparison with their radial profile is impossible as the authors considered only the neighbouring galaxies within a sphere of radius $R=1$ Mpc centred on the GRG hosts.
   Finally, we studied the location of the galaxies close to the GRGs with respect to the direction of the major axis of the source or that of the lobe expansion. With such an analysis, we tested the hypothesis whether jets expand preferentially into the direction of underdense regions. We calculated the acute angle, $\Delta \theta$, between the vector connecting the GRG host with each of the neighbouring DESI galaxies and the orientation of the source major axis, with $\Delta \theta$=0 indicating a neighbouring galaxy along the major source axis, and $\Delta \theta$=90$^{\circ}$ a galaxy which is located perpendicular
  to the source major axis. The distribution of $\Delta \theta$ appears uniform (Fig.~\ref{fig:deltapa_distribution}).  To test this, we carried out a simple bootstrapping to generate 1000 random uniform distributions of the same size as the number of GRGs of our sample, and then performed a KS test comparing our distribution with each of these generated samples, yielding a median p-value of 0.6.
  
  Based on a sample of 19 GRGs in the redshift range 0.05-0.15, \citet{Malarecki2015} found that GRG hosts live in overdense environments and that GRG lobes are shorter on the side that has a higher concentration of galaxies. Our analysis of the environment suggests that GRGs have a tendency to reside in underdense regions, even though a similar results is found for RGs as well. We found that the environment of GRG hosts on larger scales than 1-2 Mpc is not related to the orientation of the source major axis. As a consequence, only the inhomogeneities in the surrounding medium within $\sim$ 1 Mpc might play a major role in the radio galaxy evolution. Previous studies have shown that asymmetries in radio morphology (e.g., length of the lobes) can be attributed to a density gradient of the external medium \citep{Konar2008, Safouris2009, Subrahmanyan2008, Malarecki2015, Machalski2011}. 
  
\begin{figure}
        \centering
        \includegraphics[width= 0.5\textwidth]{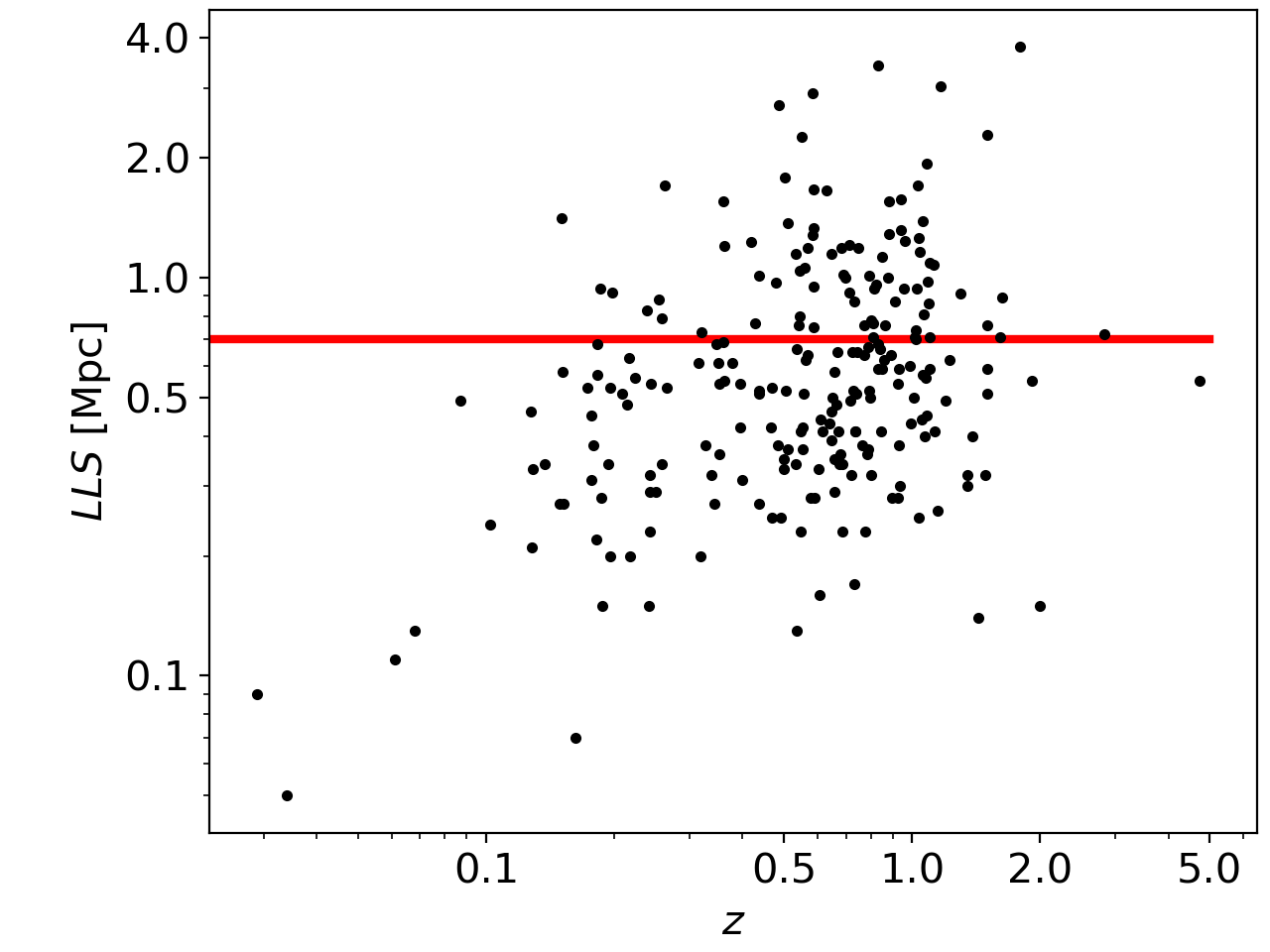}
        \caption{\textit{LLS-z} scatter plot in our sample which includes both GRGs and RGs.
        The two variables do not show any decreasing trend up to $z=2$. The red line marks the selected threshold of 0.7 Mpc.}
        \label{fig:lls-z}
  \end{figure}

  Some former studies have suggested an evolution of the linear sizes of RGs with redshift, which might be explained by the redshift evolution of the intergalactic medium \citep{Kapahi1989, Machalski2007, Onah2018}. However, \cite{Brueggen2021} found no dependence of the median linear size or the median radio luminosity on the redshift and hence no evidence for cosmological evolution of the population of GRGs. In our sample, the linear sizes of RGs do not show a decrease up to $z=2$, even though the scatter is large. At higher redshifts the sample is too small to establish a trend (Fig.~\ref{fig:lls-z}). Due to the sensitivity limitations we are not able to detect extremely large sources, especially those with low radio powers at higher redshift. Moreover, $\sim 80 \%$ of the BLDF-GRG have a redshift larger than 0.5, favouring the idea of GRGs being rare, so they are mainly detected at high redshift. We find that the volume number density of GRGs is $\rm n_{\rm grg}(\textit{z}_{\rm max}=1.8)$ = 1.00 (100 Mpc)$^{-3}$ (we excluded the only GRGs with $z>2$) which is smaller than the estimate predicted by Oei et al, in prep. who used a sample of 525 GRGs at relative low redshift.
  
  \section{Conclusions}
  \label{sec:conclusion}
   
   In this paper we carried out a detailed search for GRGs in the  Bo\"otes LOFAR deep field at 150 MHz \citep{Tasse2021} where we identified 74 GRGs with a linear size larger than 0.7 Mpc. The highest redshift GRG in our sample is a spectroscopically confirmed quasar at $z=2.84$. 
   
   We studied the properties of the host galaxies using deep optical and infrared survey data such as SDSS, DESI and WISE. We  cross-matched our GRGs in the LOFAR LBA and NVSS images and investigated their integrated spectral index. Finally, the DESI DR9 photometric redshifts enabled us to inspect the environment in which GRGs reside. The main results can be summarised as follows:
   
   \begin{itemize}
   
    \item  The GRG surface number density based on our sample is 2.8 GRGs deg$^{-2}$, which is higher than the estimates previously reported in the literature \citep[e.g.,][]{Delhaize2021, Brueggen2021}.
   
   \item The P-D diagram shows a lack of GRGs with large linear sizes ($>$ 2 Mpc) and large radio powers ($ P_{150}>10^{27}$ W Hz$^{-1}$), suggesting an evolution of the radio luminosity of such sources. In particular, the adiabatic expansion and radiative losses play a role. On the other hand, the high sensitivity of the Bo\"otes LOFAR Deep Field allowed us to detect GRGs at smaller radio powers ($10^{24} \rm W ~ Hz^{-1}$) at 150 MHz (or $3\cdot 10^{23} \rm W ~ Hz^{-1}$ at 1.4 GHz for $\alpha$=0.7). 
   
   \item The integrated spectral index is independent of the linear size of the GRG. In our analysis we find spectral indices of $\alpha$ = 0.7-0.8. However, this result may be biased by the dominance of hotspots that usually have flatter spectra.
   
   \item Most GRG hosts are LERGs, showing that FRII radio galaxies can be produced by a (currently) low accretion rate in agreement with \citet{Mingo2022}. In our sample, the optical hosts of the GRG and RG populations have a similar stellar mass distribution. In particular, both RGs and GRGs show high stellar masses ($> 10^{10.5} M_{\odot}$). The GRG hosts can be either quiescent galaxies with SFR $< 1  \rm M_{\odot} ~ yr^{-1}$ or galaxies with a moderate SFR around 10-100 $ \rm M_{\odot} ~ yr^{-1}$.
   
   \item Based on the number density of galaxies with DESI DR9 photometric redshifts, we found no significant differences between the environment density of GRG and RGs nor did we find the sources major axes to be oriented preferentially toward lower-density sectors.
   
    \item We tested whether the linear size of GRGs (\textit{LLS}$>$0.7 Mpc) and smaller RGs (\textit{LLS}$<$0.7 Mpc) have a similar distribution. We found a good agreement with the exponential and generalized Pareto distributions in both populations, even though the different scale parameters suggest that we might not probe the same distribution. The scale parameter in the BLDF-GRG sample is below the minimum size required to classify a radio galaxy as giant showing that in this work we study the extreme population of RGs.
   
   \end{itemize}
   
   In the future, we will expand our work to the two other LOFAR deep fields (ELAIS-N1 and Lockman Hole) and eventually LoTSS DR2 \citep{DR2}. Moreover, modelling the spectra from high-resolution images might help to reconstruct the evolution of AGN activity of GRGs as well as constrain the properties of the environment around these peculiar objects.

\section*{Acknowledgements}
{\small We thank the referee for their careful reading of the paper and constructive comments.
We thank L. Rudnick for preparing a low-resolution image of the LoTSS
Bootes Deep Field, which was filtered to enhance large-scale diffuse
structures. 

This work is funded by the Deutsche Forschungsgemeinschaft (DFG, German Research Foundation) under Germany's Excellence Strategy -- EXC 2121 ``Quantum Universe'' --  390833306 as well as grant DFG BR2026/27.

HA has benefited from grant CIIC\,138/2022 of Universidad de Guanajuato, Mexico. 
IP acknowledges support from INAF under the SKA/CTA PRIN “FORECaST” and the PRIN MAIN STREAM “SAuROS” projects.

LOFAR \citep{LOFAR2013} is the Low Frequency Array designed and constructed by ASTRON. It has observing, data processing, and data storage facilities in several countries, which are owned by various parties (each with their own funding sources), and that are collectively operated by the ILT foundation under a joint scientific policy. The ILT resources have benefited from the following recent major funding sources: CNRS-INSU, Observatoire de Paris and Université d'Orléans, France; BMBF, MIWF-NRW, MPG, Germany; Science Foundation Ireland (SFI), Department of Business, Enterprise and Innovation (DBEI), Ireland; NWO, The Netherlands; The Science and Technology Facilities Council, UK; Ministry of Science and Higher Education, Poland; The Istituto Nazionale di Astrofisica (INAF), Italy.
This research made use of the Dutch national e-infrastructure with support of the SURF Cooperative (e-infra 180169) and the LOFAR e-infra group. The Jülich LOFAR Long Term Archive and the German LOFAR network are both coordinated and operated by the Jülich Supercomputing Centre (JSC), and computing resources on the supercomputer JUWELS at JSC were provided by the Gauss Centre for Supercomputing e.V. (grant CHTB00) through the John von Neumann Institute for Computing (NIC).
This research made use of the University of Hertfordshire high-performance computing facility and the LOFAR-UK computing facility located at the University of Hertfordshire and supported by STFC [ST/P000096/1], and of the Italian LOFAR IT computing infrastructure supported and operated by INAF, and by the Physics Department of Turin university (under an agreement with Consorzio Interuniversitario per la Fisica Spaziale) at the C3S Supercomputing Centre, Italy.
}

\section*{Data Availability}
The datasets used for the analysis of this paper are publicly available at \href{https://lofar-surveys.org/deepfields.html}{https://lofar-surveys.org/deepfields.html}. Table~\ref{tab:list} will be made available at the CDSn and through the VizieR service (https://vizier.cds.unistra.fr).


\newpage
\clearpage

\bibliographystyle{mnras}
\bibliography{ref}

\newpage
\clearpage

\appendix

\setcounter{figure}{0}
\renewcommand{\thefigure}{A\arabic{figure}}
\setcounter{table}{0}
\renewcommand{\thetable}{A\arabic{table}}
\section*{Appendix A: Notes on individual sources}
\label{appendix}

J1419+3337 has two elongated straight and symmetric relic-type lobes. The compact source NE of the N end of the N lobe is DESI J214.7968+33.6483, $r$=14.18, and unrelated to the GRG.

J1421+3431 has a radio core slightly extended towards two faint, diffuse outer lobes.

J1421+3521  is a rather symmetric classical FR\,II source along PA $\sim100^{\circ}$. The E lobe lies at the S end of an unrelated remnant-type radio galaxy hosted by 2MASX\,J14220065+3523086 with \textit{LAS}$\sim3.3 \rm '$, with its N lobe outside of our cutout.

J1422+3320 is a very straight FR\,II type source with relatively narrow lobes showing emission 
decreasing in surface brightness from the hotspots right to the host galaxy. Also
 displayed in \cite{Williams2016}.

J1422+3255  No radio core is detected, and the source may also be hosted by DESI\,J215.6873+32.9360, $r$=24.69, with $z_{ph}=1.020$, although this would not 
alter its \textit{LLS}.

J1423+3600  is a restarted or double-double radio galaxy with an inner double of $\sim$24$\rm ''$ size and both inner and outer double having stronger N lobes. 

J1423+3403  shows some evidence for being restarted with two faint inner lobes $\sim$24 $\rm ''$ apart and outer lobes being of remnant type. About 32$\rm ''$ NNE of the host a compact radio source (SDSS\,J142337.34+340330.2 at $z_{\rm ph}\sim$0.42) is superposed on the NNE lobe.

J1423+3302: The BLDF image is slightly affected by sidelobes, but both VLASS epoch 1 and 2 images confirm this is a restarted source, with an inner double of $\sim$27 $\rm ''$.  The ENE outer lobe is resolved out in VLASS while the WSW outer lobe shows two hotspots, though the W hotspot 
may be hosted by DESI J215.8935+33.0419, $r$=23.79, at a similar redshift as the host.

J1423+3529  has the second-largest linear size (3.4\,Mpc) in our sample. It is a restarted source
 with an inner asymmetric double of \textit{LAS}$\sim$1.6 $\rm '$. The 6$ \rm ''$ BLDF image shows rather disrupted, and generally straight outer lobes, but the LoTSS-DR2 low-resolution image leaves no doubt that the source is a physical entity.  Only the host is detected as a faint point source in VLASS.

J1423+3340 is a symmetric and straight FR\,II source with a core slightly brighter than the lobes
in BLDF. The host is only detected in CWISE, thus we estimate a redshift of at least $\sim$1.5.

J1423+3605 is a WAT-type source with radio tails due SSE and W in a poor cluster or group. An apparent loop of radio emission in BLDF W of the host is caused by extended radio emission
of a pair of galaxies at the same redshift and $\sim$11$\rm ''$ SW of the host.

J1424+3609 is a very large and straight FR\,II source with clear diffuse emission trailing back from the outer extremities of the NNE and SSW lobes towards the host. Only the NNE lobe shows a hotspot in VLASS while the end of SSW lobe is diffuse and of very low surface brightness in VLASS. A compact radio source $\sim$20$\rm ''$ SSW of the host has no optical or IR counterpart and is likely a knot of the SSW jet. The source was first noticed by us in \cite{Williams2016}.

J1424+3436 is likely a restarted radio galaxy with inner lobes of size $\sim$50$\rm ''$ but no clear
hotspots in VLASS. Beyond the NE inner lobe a collimated radio feature connects with the diffuse outer NE lobe, while beyond the SW inner lobe the radio emission connecting with the SW lobe seems bifurcated. The outer lobes show distinct radio morphologies with only the SW one showing two hotspots in VLASS. First seen by us in the 62-MHz cutouts by \cite{VanWeeren2014}, albeit with much lower sensitivity than BLDF.

J1425+3633 is a regular, straight, but slighty asymmetric FR\,II with outer lobes highly dilute
in VLASS and no hotspots. 

J1425+3506 is an FR\,II source with faint, dilute lobes. The host could also be DESI J216.2998+35.1115, r=25.00, with $z_{\rm ph}$=0.971,1.308,1.56 from \cite{Zhou2021}, \citet{Duncan2021}, \citet{Chung2014} in which case its \textit{LLS} would be $\sim$1.1\,Mpc.

J1425+3557 is a core-dominated source with very dilute remnant-type lobes extending due E and W. An extended brightening $\sim$30$\rm ''$ from the host due W has no optical or IR counterpart and may qualify it as a hybrid-morphology source (HyMORS).

J1426+3407C is a candidate FR\,II source since the SSE lobe has SDSS\,J142625.73+340657.8  as a possible host, and the NNW lobe features SDSS J142622.04+340902.0 at its S end.

J1426+3320 is an FR\,II source where both lobes bend from the E-W direction towards S, giving it a C shape. It is completely resolved out in VLASS with no hotspots.

J1426+3236 is a regular, straight and symmetric FR\,II source completely resolved out in VLASS, suggesting that the hotspots have already disappeared. Also listed in \cite{Mingo2022}.

J1426+3222 is an FR\,II source with diffuse lobes towards E and W, already shown with a lower-sensitivity image by \cite{Williams2016}. From the inner parts of each lobe faint radio plumes stretch out due N.

J1427+3625 is an asymmetric, slightly bent FR\,II source extending along PA$\sim$48$^{\circ}$, with the NE lobe featuring a curved ridge of emission stretching from the hotspot first due SE and then SW towards the host. The SW lobe only shows a short trail from the end of the lobes towards the host. This is the only source in our sample that has been reported before as a giant radio quasar by \cite{Kuzmicz2021}.

J1427+3312: Its host was detected in X-rays (CXOXB\,J142718.3+331205) by \cite{Brand2006}. The radio emission features a strong, short NW lobe and a much longer and lower surface brightness SE lobe whose exact length is somewhat uncertain, making it a hybrid morphology source.

J1427+3328C: the source with the largest angular size in our sample with very inflated and low surface brightness lobes, with the E lobe slightly flattened with major axis along PA$\sim$70$^{\circ}$ and the W lobe closer to the host and extended in the N-S direction, almost perpendicular to the major axis of the source. Several compact sources are superimposed on the lobes and it is difficult to state a precise total flux. There is no radio emission detected at the location of our proposed host 2MASX J14273770+3328081. However, it is the brightest galaxy in both the optical and MIR (unWISE) in the entire region of the radio emission. Also, its angular and linear sizes may be larger in more sensitive future radio images. 
No radio emission detected at the location of our proposed
host 2MASX~J14273770+3328081. However, it is the brightest galaxy in
both the optical and mid-infrared (unWISE) in the entire region of the radio
emission. There is in fact a second-brightest galaxy in this region
46$''$~SSW of 2MASX~J14273770+3328081, SDSS J142736.24+332726.0, $r$=18.03,
$z_{sp}$=0.231 (Kochanek et al., 2012), which does coincide with a radio
source of 150-MHz flux of 0.37-mJy in the LoTSS Bootes Deep Field.
which is also closer to lying mid-way between the flux-weighted centers of each lobe. 
If this were the true host, its $LLS$ would be 2.0~Mpc.

J1427+3255C is a straight, symmetric classical FR\,II source along PA$\sim$135$^{\circ}$. 
The 
SE lobe shows a diffuse radio tail (or plume) due N from the termination point, 
while the NW lobe is fainter and more compact in the BLDF image. We consider it a candidate GRG, although neither lobe has a suitable optical or IR counterpart.

J1428+3432 is a remnant-type FR\,II radio galaxy with a clear core but very diffuse, non-collinear lobes due NE and S.

J1428+3631 is a classical FR\,II source already shown by \cite{Williams2016} with lobe
emission almost all the way from the hotspots (clearly seen in VLASS) and the
radio core (faint, but detected in VLASS). There is another, smaller and unrelated
FR\,II source $\sim$1.6$\rm '$ SW of the host.

J1428+3446 is a core-dominated FR\,II type source with rather straight and very low surface
brightness lobes along PA$\sim$150$^{\circ}$. Background sources are superposed due W of the S lobe, as well as to the SW and NE of the N lobe. 

J1428+3525 is a remnant-type radio galaxy with a clear core, a shorter and stronger W lobe, and a longer and fainter E lobe. Both lobes are of very low surface brightness and are aligned with the host, suggesting that a few starlike optical objects in these lobes are superposed accidentally.

J1429+3326: the source features an unusual emission spur or plume emanating from the SW
lobe, close to the host, in the NW direction and perpendicular to the main source axis, also seen in Fig.\,A1 of \cite{Williams2016} but too faint to be seen in the 62-MHz image by \cite{VanWeeren2014}. The source is completely resolved out in VLASS. Such features have been seen in both lobes of X-shaped sources like in
PKS\,2014$-$55 \cite{Cotton2020} but in J1429+3326 we see it only in one lobe.

J1429+3355 is a straight and slightly asymmetric FR\,II radio galaxy with the shorter NW lobe
being strong and featuring a hotspot in VLASS, while the SE lobe is fainter in BLDF
and undetected in VLASS suggesting the absence of a hotspot. This source was first noticed by us in a 325-MHz VLA image kindly supplied by R.\ Coppejans \citep{Coppejans2015}, but is also displayed in \cite{Williams2016}.

J1429+3356 is a core-dominated FR\,II source with a collimated (jet?) feature half-way from 
the core to the NW lobe. This feature is very faint in VLASS which also detects the 
end of the NW lobe as a diffuse feature (no hotspot), while the SE lobe is completely undetected. The BLDF image shows some background sources $\sim$30$\rm ''$ SW of the SE lobe and $\sim$30$\rm ''$ N of the end of the NW lobe.  This source was first noticed by us in a 325-MHz VLA image kindly supplied by R.\ Coppejans \citep{Coppejans2015}.

J1429+3717 is a remnant-type radio galaxy which shows a clear core, a short E lobe and a much larger SW lobe which has an unrelated strong FR\,II source superposed. The latter does not permit us to measure a reliable size for the diffuse source which is completely resolved out in VLASS.

J1429+3230C: this source, supposedly a large FR\,II with major axis along PA$\sim$165$^{\circ}$,
is somewhat speculative, since the faint S lobe has optical and IR objects superposed.
However, the N lobe is elongated along the source major axis and does not feature suitable optical or IR counterparts. The suggestion of a $\sim$2.2$\rm '$ wide double source straddling the host along PA$\sim$100$^{\circ}$ is due to two sources aligned with the host but with clear counterparts.

J1430+3519 is a symmetric and straight FR\,II radio galaxy with strong, but very diffuse remnant-type lobes. Unusual features are two compact sources, one at the SW edge of the SE lobe and another close to the middle of the NW lobe, neither one having an obvious optical or
IR counterpart.  This source has the steepest radio spectrum in our sample and was first seen by us in \cite{Williams2016}.

J1430+3322C: this faint source features an inner FR\,I-type structure of $\sim$32$\rm ''$ E-W
extent, with the core faintly detected in VLASS, including a slight E-W extent. The stronger W lobe is closer to the host than the more detached E lobe located $\sim$1.8$\rm '$ from the host. The source may well extend further into a WAT-type source in future, more sensitive images.

J1431+3345 is probably a good example of an FR\,I source with continuous jets from the 
host 
towards both lobes which seem to end in a strong bend, perhaps seen in project in the SW lobe. Due to the presence of these outer lobes we classify it as an FR\,I/II. While the jets appear rather straight for the first arcmin from the host, their PA differs by about 7$^{\circ}$.
The host is the brightest cluster galaxy in cluster WHL J143103.5+334542 \citep{Wen2012}. This source was first noticed by us in a 325-MHz VLA image kindly supplied by R.\ Coppejans \citep{Coppejans2015}, but is also seen in
\cite{Williams2016}.

J1431+3535C is a relatively straight FR\,II source with lobes of different morphology. Both lobes have optical/IR objects superposed, but in our opinion these are less likely to explain the radio structure. There is a circularly extended source immediately NW of the host, perpendicular to the major source axis, which has no optical/IR counterpart. The strong extended radio source  $\sim$1$\rm '$ W of the N end of the source is the bright foreground starburst galaxy MCG\,+06-32-056 with a dust lane. The r=18.1 mag low surface brightness galaxy is superposed on the SW lobe, but its blue color and low z$_{phot}$ suggest it is a starforming galaxy in the foreground.

J1431+3440 is a straight and symmetric FR\,II source with diffuse lobes, each of them showing
plumes due W. The source is difficult to recognize due to various compact sources due W and NW of the host.   

J1431+3427 is a straight and symmetric FR\,II source with a faint radio core.  This source was first noticed by us in a 325-MHz VLA image kindly supplied by R.\ Coppejans 
\citep{Coppejans2015}, but is also displayed in \citet{Williams2016}.

J1432+3411 is a symmetric, though slightly bent FR\,II source with no detected hotspots in 
VLASS. The BLDF image suggests the jets feeding the lobes to have a difference
in PA of $\sim$7$^{\circ}$.  This source was first noticed by us in a 325-MHz VLA 
image kindly supplied by R.\ Coppejans \citep{Coppejans2015}, but is also 
displayed in \citet{Williams2016}.

J1432+3328 is a rather asymmetric, though straight, FR\,II with an armlength (or lobelength) ratio of 2.8. The source is listed in \cite{Mingo2022}.

J1432+3545 is a very extended WAT-type source hosted by the brightest galaxy of cluster
WHL J143233.9+354540 (aka GMBCG J218.14135+35.76114 and MaxBCG J218.14136+35.76113).
The angular size we list may well be exceeded by future more sensitive observations.

J1432+3220 is a  symmetric FR\,II source with faint hotspots and a faint core detected in VLASS. This source was first noticed by us in a 325-MHz VLA image kindly supplied by R.\ Coppejans \citep{Coppejans2015} and is also seen in \citet{Williams2016}.

J1432+3154C is a rather faint FR\,II source with diffuse lobes and a core symmetrically
placed between these lobes. We consider it a candidate because each lobe has optical/IR objects which may explain the lobes as separate sources. No part of the source is detected in VLASS.

J1432+3647 is an FR\,II source with detached, diffuse, low surface brightness lobes bent at an angle of $\sim$25$^{\circ}$ with respect to the host. The NE lobe appears bifurcated. Only the radio nucleus is detected in VLASS. 

J1433+3220 is a possible restarted FR\,II source with an inner double of $\sim$38$\rm ''$ along PA$\sim$104$^{\circ}$, very close to the PA connecting the diffuse, low surface brightness outer lobes. Since these lobes are large, some optical/IR objects are seen superposed, but they are unlikely to explain each lobe as a separate source. The host itself is undetected in both BLDF and VLASS, which does not detect any part of the source.

J1433+3328 is a core-dominated straight FR\,II source oriented E-W and hosted by a QSO at $z_{\rm sp}$=1.609. While the W lobe has a typical FR\,II morphology being brightest at the end in BLDF with a barely detected hotspot in VLASS, the E lobe is brightest about half way down the lobe, and is undetected in VLASS. It is thus a possible hybrid-morphology source.

J1433+3450 is a classical, slightly bent FR\,II  source with a faint core detected in both BLDF and VLASS. While our proposed host coincides with that radio core, it differs from that proposed by \cite{Morabito2017} which is off the main radio ridge. The hotspots are strong in VLASS, albeit
with much  shorter trails towards the host than those in BLDF. This source was first noticed by us in \cite{VanWeeren2014} and is also seen in \cite{Williams2016}. This is the source with the highest radio luminosity in our sample.

J1433+3145C is a low surface brightness diffuse radio galaxy without a clear radio core. The LoTSS DR2 low-resolution image suggests a WAT-type source whose host may also be located further South, e.g. SDSS\,J143320.32+314544.0 with $z_{\rm ph}\sim$0.4 which would make its \textit{LLS} about 0.77\,Mpc, but still large enough to be a GRG. The strongest compact 150-MHz source at the S end of the emission region, ILT~J143320.57+314529.1 of $\sim$1\,mJy,  does not have any optical/IR counterpart.

J1434+3428 is a complex, slightly bent FR\,II source with several radio knots along the E lobe. Both lobes feature hotspots in BLDF which are also faintly detected in VLASS, but which are located due N of the symmetry axis of the lobes.  The host of this source is still unclear. We have chosen the brighter one, but $\sim$19$\rm ''$ E of it there is  WISEA J143430.65+342757.0 with more typical AGN colors, undetected in the optical in DESI\,DR9, but found to be a member of cluster ISCS\,J1434.5+3427 with $z_{\rm sp}$=1.240 from \citet{Alberts2016}, also detected as X-ray source 2CXO\,J143430.6+342757, and as FIRST J143430.6+342757 in the radio. Adopting the latter host this source would be only slightly larger and more radio luminous, as well as a 
candidate restarted source.  The source was first noticed by us in a 325-MHz  VLA image kindly supplied by R.\ Coppejans \citep{Coppejans2015}.

J1434+3214C has a clearly detected inner source of $\sim$1.0$\rm '$ extent that is accompanied by
fainter lobes more or less symmetrically placed on the E and W side of the latter, making it more of an FR\,I rather than FR\,II type source.
The source fades away below noise at \textit{LAS}$\sim$3.7$\rm '$ but may continue further in more sensitive images.

J1434+3648 is an almost straight FR\,II source with low-surface brightness or remnant-type lobes along PA$\sim$140$^{\circ}$. Some optical/IR objects in the area of the lobes seem unlikely to explain them individually, so we consider the
source as genuine. 

J1434+3328C  is a very large diffuse source of oval shape with major axis near PA$\sim$160$^{\circ}$ with a low surface brightness extension from the end of the N lobe towards SW. The N lobe is much brighter and of FR\,II type, while the S lobe is fainter and of FR\,I type, thus it is a candidate hybrid morphology source.  The innermost  source shows a compact radio knot $\sim$6$\rm ''$S of the host, also seen in VLASS  and \citet{Coppejans2015}. An additional diffuse source is seen $\sim$16$\rm ''$ from the host with no optical/IR counterpart. The host itself is a broadline QSO at $z_{\rm sp}$=0.19756 (from SDSS) and is apparently located in a compact triple of galaxies. However, while the 2nd-brightest member of that triple, SDSS\,J143444.88+332817.3 $\sim$6$\rm ''$WSW of the host has a $z_{\rm ph}$ compatible with the host, the 3rd-brightest member (SDSS\,J143445.33+332823.5)  $\sim$3$\rm ''$N of the host is a galaxy at $z_{\rm sp}$=0.24565, so apparently in the background. The host was also detected in X-rays as 1RXS\,J143445.8+332814 and more recently with Chandra \citep{Masini2020} and \cite{Kostrzewa-Rutkowska2018} report the host to have featured a Gaia transient. The host had been proposed as a blazar candidate by \cite{dabrusco2019}. It is thus possible that the source is oriented at a large angle with respect to the plane of the sky and is intrinsically much larger. The peak of the emission of the N lobe in the
full-res. Deep Field coincides with SDSS J143442.62+333015.8 with z$_{phot}$
indistinguishable from the QSO host's z$_{spec}$, so that galaxy could be
the host of a WAT-like source that consititute most of the NNW lobe
including the circular wings running to SSE and SSW, making the inner
source only about 3.4$\rm '$ (LLS = 0.67 Mpc), and N WAT-type source would
have 3$\rm '$ or LLS=0.59 Mpc.
           
J1435+3404 is a rather faint Z-shaped source with a clear radio nucleus and evidence for precessing jets or lobes starting in the E-W direction and gradually turning S in the E lobe and N in the W lobe.

J1435+3547: the BLDF image shows this to be a straight and symmetric restarted FR\,II radio galaxy with inner lobes of \textit{LAS}$\sim$32$\rm ''$. Except for very faint hotspots it is not detected in VLASS, suggesting both inner and outer lobes having aged substantially, which is consistent with the fact that it is the source with the second-steepest radio spectrum in our sample.

J1436+3416 is a straight, restarted FR\,II with asymmetric inner lobes of \textit{LAS}$\sim$1.1$\rm '$. The inner lobes have well-detected hotspots in VLASS but no detection of the host. The hotspot of the inner NE lobe is displaced significantly from the galaxy SDSS\,J143654.42+341726.9 near the middle of that lobe and we regard this galaxy as accidentally superposed. The outer lobes are undetected in VLASS. Our proposed host is undetected in WISEA, and closer to the center between the inner hotspots there is WISEA J143653.20+341657.1 with 
AGN-like WISE colors and undetected in DESI DR9, but listed with $z_{\rm ph}$=1.54 in \citet{Duncan2021}, which would imply an \textit{LLS} of $\sim$1.88\,Mpc. Some 52$ \rm ''$ NW of the host the very bright foreground galaxy MCG +06-32-076 appears as a strong
radio source.

J1437+3650 is a slightly asymmetric remnant-type FR\,II source with low surface brightness lobes along PA$\sim$140$^{\circ}$. Part of the farther SE
lobe may be caused by emission from superposed objects.

J1437+3233 is a WAT-type source with indications of the far WSW lobe showing a bend of $\sim$150$^{\circ}$, perhaps due to projection. The host is too faint to recognize any cluster around it on optical images, but the unWISE image \citep{Lang2014} suggests a concentration of objects around the source.

J1438+3445C is a slightly bent FR\,II source with very detached lobes. Only the stronger SE lobe, which has no optical/IR counterpart, shows a faint indication of a hotspot in VLASS.

J1438+3526 is a straight, core-dominated FR\,II source with a stronger NE lobe terminating in what appears as a hotspot, which is undetected in VLASS. The SW lobe is of very low surface brightness and suggests a bend due SE near its end.

J1438+3355: this is the third-largest source in angular size but the one of largest linear size (3.8\,Mpc) in our sample. The outer lobes are very detached from the host, but both show the characteristic (though short) trail of radio emission from the far end of the lobes towards the host. The lobe length ratio is only 1.4 (NE/SW lobe) and both lobes are aligned with the host to within 1.5$^{\circ}$. The source has a radio luminosity in the upper quartile and a rather steep spectrum ($\alpha>1.17$).

J1438+3539 is a rather asymmetric source with the southern more compact lobe reaching twice as far from the host than the northern, more diffuse lobe. A faint hotspot is seen in the S lobe in VLASS, separated from two WISE objects E of it. This, together with the fact that the S lobe has a faint, low surface brightness spur due NE, suggests that this is in fact the S lobe of a large source. Also, the lobe orientations with respect to the host are aligned to within 2$^{\circ}$.

J1439+3251C: this candidate radio galaxy features two symmetric, low surface brightness lobes along PA$\sim$54$^{\circ}$, and we include in its size a compact source at the end of the NE lobe which we interpret as a hotspot since it does not show any optical/IR counterpart.This hotspot is the only part of the source detected in VLASS. We chose the host as it features an apparent radio core in BLDF, while a similarly bright galaxy due E (SDSS\,J143903.20+325135.9) does not.

J1439+3254 is a straight, symmetric FR\,II source of high radio luminosity. VLASS shows diffuse lobes with indications of two hotspots in the stronger N lobe and a very faint one in the S lobe.  This source was first noticed by us in a 325-MHz VLA image kindly supplied by R.\ Coppejans \citep{Coppejans2015}.

J1439+3221C is a core-dominated remnant-type source of intermediate FR type. The N lobe shows some wiggling before it reaches its maximum brightness at its end, though a hotspot is not detected in VLASS. The S lobe shows continuous very low surface brightness emission in the low-resolution LoTSS DR2 image before it seems to fade away below noise $\sim$2.5$\rm '$ S of the host, but may in fact continue for another 1.3$ \rm '$ due S.

J1440+3211 is one of the angularly largest sources in our sample, this is an FR\,II source with very detached lobes, both of them showing radio trails from their end towards the host, which both lobes being oriented parallel to within less than 1$^{\circ}$. The host is barely detected in BLDF, and undetected in VLASS. The N lobe shows a prominent hotspot in VLASS while in the S lobe none is detected.

J1440+3348: this source is similar to the protypical FR\,I sources with precessing jets (like 3C\,31 or 3C\,449), though the jets are below the detection limit of VLASS.

J1442+3457 is a rather straight FR\,II type source with a very detached NW lobe and a more diffuse SE lobe, both of low surface brightness. The more compact source $\sim$30$\rm ''$ N of the N lobe has a separate host and is thus unrelated.

J1442+3605: this FR\,II type source has one of the highest radio luminosities in our sample and VLASS resolved the lobes into broad, almost circular emission regions of almost constant surface brightness, albeit with hotspots embedded in them. These hotspots are not located at the outer edges of the lobes, such that the distance between them is 1.5$ \rm '$ compared to 1.84$ \rm '$ for the total projected size of the source.

J1442+3243 is a straight, asymmetric FR\,II type source with strong lobes in VLASS showing constant surface brightness over an oval region in which a hotspot (or at least a "spine") is embedded. The exact size of this source is difficult to measure on its E side where it overlaps with the NE lobe of J1442+3242.  This source was first noticed by us in a GMRT 150-MHz image by \cite{Williams2013}.

J1442+3242: the host of this FR\,II source is only $\sim$18$\rm ''$ SE of the E hotspot of J1442+3243, thus its diffuse NNE lobe overlaps with the E lobe of J1442+3243. Only the stronger SSW lobe shows a well-detected hotspot in VLASS surrounded by a faint envelope of diffuse emission. Future, improved VLASS images may confirm a possible \textit{LAS} of up to 2.0$ \rm '$.

J1444+3445C is one of the sources with the lowest radio luminosity in our sample. It has an FR\,I type inner region with two very low surface brightness lobes due W and ESE with a bending angle of $\sim$25$^{\circ}$ between them. Some optical/IR objects are superposed on the outer lobes which we consider unlikely to account for the lobes, which are entirely undetected in VLASS.

J1444+3348: this source has an uncertain radio morphology and features a faint (if any) radio core, a strong but diffuse W lobe close to the host and a longer, more collimated E lobe which bends due N close to its end. 

J1444+3444 is a rather straight FR\,II source oriented along PA$\sim$157$^{\circ}$ and is difficult to recognize due to a compact source superposed $\sim$18$\rm ''$ WNW of the host.  It features a continuous SSE lobe of moderate surface brightness that curves due W before its end, and a more detached and patchy NNW lobe. No part of this source (except the 
superposed one) is detected in VLASS.

\begin{sidewaystable*}
\centering
\caption{Col.~(1), name of the giant radio galaxy. The superscript refers to the reference of the catalogue that reported the source: 1-\citet{Coppejans2015}, 2-\citet{Williams2016}, 3-\citet{VanWeeren2014}, 4-\citet{Kuzmicz2021}, 5-\citet{Williams2013}, 6-\citet{Mingo2022}, 7-\citet{Masini2021}. Col.~(2) and Col.~(3), right ascension and declination (J2000) of the host galaxy in degrees. Col.~(4), largest angular size. Col.~(5), classification of GRGs according to the Fanaroff–Riley classification. Col.~(6), redshift of the host galaxy. Col.~(7), redshift error when available. Errors of the spectroscopic redshifts are not reported since they are generally more accurate than the precision we can achieve on the linear sizes and luminosities. Col.~(8), type of the redshift: p for photometric, s for spectroscopic and e if estimated. Col.~(9), references for the redshift. These are either spectroscopic measurements, such as 3-SDSS\citep{sdssdr12},  8-LAMOST\citep{lamost}, 9-\citet{Hucra2012}, 10-\citet{Kochanek2012}, or photometric
estimates: 1-DESI \citep{Zou2019, Zhou2021}, 2-\citet{Duncan2021}, 4-Pan-STARRS \citep{Beck2021}, 5-\citet{Chung2014}, 6-\citet{Bilicki2016}, 7-\citet{Brescia2014}. Col.~(10), name of the host galaxy. Col.~(11) type of host galaxy: galaxy (G) or QSO (Q) or candidate quasars (Qc). Col.~(12), largest linear size. Col.~(13), magnitude of the host galaxy in the r-band if available from the DESI DR9 photometric catalogue; the label W1 and W2 indicates that the magnitude is taken from either WISEA or CWISE catalogue. Col.~(14), total flux at 150 MHz from the full resolution 6$"$ image. An asterisk is appended to the flux estimate of the source if the flux integration is computed by considering all the pixels belonging to the region of the radio emission. Col.~(15) total flux 1$\sigma$ error from the 6$"$ image. Col.~(16), log10 of power at 150 MHz. Col.~(17) spectral index $\alpha$ measured by using NVSS (1.4GHz), LOFAR HBA (150MHz) and LOFAR LBA (50MHz) data. Col.~(17), spectral index error. }
    \begin{tabular}{p{2cm}p{1cm}p{1cm}p{0.3cm}p{0.3cm}p{0.5cm}p{0.5cm}p{0.4cm}p{1cm}p{4cm}p{0.4cm}p{0.6cm}p{1.3cm}p{0.5cm}p{0.4cm}p{0.6cm}R{0.9cm}p{0.5cm}}
            \hline
                     (1) & (2) & (3) & (4) & (5) & (6) & (7) & (8) & (9) & (10) & (11) & (12) & (13) & (14) & (15) & (16) & (17) & (18)  \\
                    Name & RA$_J$ & ${\rm Dec}_J$ & \textit{LAS} & FR & $z$  & $\Delta z$ & ztype & ref & Hostname & type & \textit{LLS} & mag & $S_{150}$ &  $\sigma_S$ & $\log_{10}P_{150}$ & $\alpha$ & $\Delta \alpha$  \\ 
                         LBDF-GRG & $^{\circ}$ & $^{\circ}$ & ($'$) &type &  & & &  & & & (Mpc) & (mag) & (mJy) & & (W/Hz) & &  \\ \hline
                
        J1419+3337 & 214.8027 & 33.6328 & 1.8 & II & 1.09 & 0.16 & p & 1,4 & DESI J214.8027+33.6329 & ? & 0.9 & 23.41 r & 9.6 & 1.4 & 25.7 & ~ & ~ \\ 
        J1421+3431 & 215.4634 & 34.5289 & 1.5 & II & 1.02 & 0.25 & p & 1 & DESI J215.4634+34.5289 & G & 0.7 & 24.59 r & 4.1 & 0.8 & 25.2 & 0.66 & 0.19 \\ 
        J1421+3521 & 215.4909 & 35.3570 & 2.2 & II & 1.1 &  & e & - & CWISE J142157.80+352125.3 & ? & 1.1 & 17.31 W2 & 58.6 & 5.9 & 26.5 & 0.93 & 0.05   \\ 
        J1422+3320 & 215.6373 & 33.3354 & 2.8 & II & 0.94 & 0.11 & p & 1,4 & DESI J215.6373+33.3353 & G & 1.3 & 23.83 r & 145 & 15 & 26.7 & ~ & ~ \\ 
        J1422+3255 & 215.6913 & 32.9322 & 1.7 & II & 1.07 & 0.23 & p & 1 & DESI J215.6913+32.9322 & G & 0.8 & 24.17 r & 53.8 & 5.5 & 26.4 & ~ & ~\\ 
        J1423+3600 & 215.7846 & 36.0052 & 2.0 & II & 1.09 & 0.21 & p & 1 & DESI J215.7846+36.0053 & G & 1.0 & 24.16 r & 55.6 & 5.7 & 26.5 & 0.94 & 0.05 \\ 
        J1423+3403 & 215.9003 & 34.0506 & 2.7 & I/II & 0.561 &  & s & 3 & SDSS J142336.06+340302.1 & G & 1.1 & 20.17 r & 9.9 & 1.4 & 25.0 & 0.86 & 0.12 \\ 
        J1423+3302 & 215.9105 & 33.0484 & 1.9 & II & 1.03 & 0.19 & p & 1,4 & DESI J215.9106+33.0483 & G & 0.9 & 23.82 r & 258 & 26 & 27.1 & ~ \\ 
        J1423+3529 & 215.9201 & 35.4888 & 7.5 & II & 0.83 & 0.05 & p & 1,3,4 & SDSS J142340.83+352919.7 & G & 3.4 & 21.91 r & 20.4 & 3.0 & 25.7 & 1.10 & 0.12 \\ 
        J1423+3340 & 215.9259 & 33.6693 & 1.5 & II & 1.5 &  & e & - & CWISE J142342.22+334009.6 & ? & 0.8 & 18.05 W1 & 3.6 & 0.9 & 25.6 & 0.48 & 0.14 \\ 
        J1423+3605 & 215.9486 & 36.0946 & 2.8 & I & 0.36 & 0.02 & p & 1,3,6 & SDSS J142347.66+360540.5 & G & 0.8 & 18.73 r & 40.7 & 4.2 & 25.2 & ~ \\ 
        J1424+3609$^{2}$ & 216.0397 & 36.1547 & 7.3 & II & 0.58 & 0.07 & p & 1,3,4 & SDSS J142409.52+360917.0 & G & 2.9 & 20.89 r & 191 & 19.3 & 26.4 & 0.74 & 0.04\\ 
        J1424+3436$^{2,3}$ & 216.1358 & 34.6127 & 4.2 & I/II & 0.586 &  & s & 3 & SDSS J142432.58+343645.8 & G & 1.7 & 20.14 r & 486 & 49 & 26.8 & 0.85 & 0.04 \\ 
        J1425+3633 & 216.2942 & 36.5612 & 2.3 & II & 0.70 & 0.08 & p & 1,3,4,7 & SDSS J142510.60+363340.3 & G & 1.0 & 21.13 r & 217 & 22 & 26.6 & 0.89 & 0.04\\ 
        J1425+3506$^6$ & 216.2995 & 35.1045 & 2.1 & II & 0.71 & 0.05 & p & 1,2,3,4 & SDSS J142511.88+350616.1 & G & 0.9 & 22.09 r & 14.3 & 1.7 & 25.4 & 0.94 & 0.12\\ 
        J1425+3557 & 216.3287 & 35.9574 & 3.7 & II & 0.419 &  & s & 3 & SDSS 142518.87+355726.7 & G & 1.2 & 18.54 r & 14.6 & 1.7 & 24.9 & ~ \\ 
        J1426+3407C & 216.6021 & 34.1285 & 2.7 & II & 0.75 & 0.14 & p & 1,2,3,4 & SDSS J142624.49+340742.6 & G & 1.2 & 21.96 r & 42.8 & 4.5 & 26.0 & 1.02 & 0.06 \\ 
        J1426+3320$^6$ & 216.6026 & 33.3373 & 2.1 & II & 0.82 & 0.15 & p & 1,2 & DESI J216.6026+33.3373 & G & 1.0 & 23.66 r & 62.2 & 6.3 & 26.2 & 1.19 & 0.05 \\ 
        J1426+3236$^6$ & 216.6109 & 32.6024 & 2.0 & II & 0.96 & 0.11 & p & 1,2 & DESI J216.6109+32.6024 & G & 0.9 & 24.05 r & 46.8 & 4.8 & 26.2 & ~ \\ 
        J1426+3222$^2$ & 216.7126 & 32.3717 & 2.8 & II & 0.68 & 0.12 & p & 1,2,3,4,5 & SDSS J142651.02+322218.0 & G & 1.2 & 21.36 r & 386 & 39 & 26.8 & 0.92 & 0.04 \\ 
        J1427+3625$^4$ & 216.8259 & 36.4170 & 2.4 & II & 0.851 & ~ & s & 3 & SDSS J142718.21+362501.1 & Q & 1.1 & 18.02 r & 200 & 20 & 26.8 & 0.72 & 0.04 \\ 
        J1427+3312 & 216.8265 & 33.2016 & 2.1 & I/II & 0.93 & 0.28 & p & 1,2,4,11 & DESI J216.8265+33.2016 & G & 1.3 & 22.89 r & 18.3 & 1.9 & 25.8 & 1.12 & 0.11\\ 
        J1427+3328C & 216.9070 & 33.4688 & 9.0 & II & 0.150 &  & s & 3,10 & 2MASX J14273770+3328081 & G & 1.4 & 16.66 r & 64.7* & 7.2 & 24.6 & $>$0.26 \\ 
        J1427+3255C & 216.9363 & 32.9289 & 2.7 & II & 2.18 & 1.65 & p & 2 & SSTSL2 J142744.69+325543.9 & G & 1.4 & 16.49 W2 & 7.5* & 1.0 & 26.3 & ~ \\ 
        J1428+3432 & 217.0151 & 34.5361 & 1.4 & II & 1.02 & 0.11 & p & 1,2,4 & SDSS J142803.63+343210.0 & G & 0.7 & 23.24 r & 3.9 & 0.7 & 25.2 & $>$0.30 \\ 
        J1428+3631$^2$ & 217.0284 & 36.5249 & 2.4 & II & 0.59 & 0.19 & p & 1,3,7 & SDSS J142806.81+363129.5 & G & 0.9 & 21.60 r & 602 & 60 & 26.9 & ~ \\ 
        J1428+3446 & 217.0977 & 34.7752 & 1.9 & II & 0.588 &  & s & 11 & SDSS J142823.44+344630.6 & G & 0.7 & 20.67 r & 7.5 & 1.0 & 25.0 & ~ \\ 
        J1428+3525 & 217.1499 & 35.4271 & 2.8 & II & 0.88 & 0.23 & p & 1,2,4,5 & DESI J217.1498+35.4271 & G & 1.3 & 23.16 r & 10.9 & 1.4 & 25.5 & \\ 
        J1429+3326$^{2,3}$ & 217.2997 & 33.4437 & 2.6 & II & 0.320 &  & s & 8 & SDSS J142911.92+332637.5 & G & 0.7 & 18.83 r & 270 & 27 & 25.9 & ~ \\ 
        J1429+3355$^{1,2}$ & 217.3171 & 33.9270 & 2.8 & II & 1.059 &  & s & 3 & SDSS J142916.10+335537.4 & G & 1.4 & 23.03 r & 137 & 14 & 26.8 & 0.92 & 0.04 \\ 
        J1429+3356$^{1,2,4}$ & 217.4277 & 33.9486 & 2.2 & II & 1.124 &  & s & 3 & SDSS J142942.63+335654.8 & Q & 1.1 & 18.97 r & 92.9 & 9.3 & 26.7 & 0.88 & 0.04 \\ 
        J1429+3717 & 217.4724 & 37.2914 & 2.8 & II & 0.65 & 0.06 & p & 1,3 & SDSS J142953.37+371728.9 & G & 1.1 & 21.95 r & 22.6 & 2.4 & 25.5 & ~ \\ 
         \hline

         \label{tab:list}

\end{tabular}
\end{sidewaystable*}

\begin{sidewaystable*}
\centering
    \begin{tabular}{p{2cm}p{1cm}p{1cm}p{0.3cm}p{0.3cm}p{0.5cm}p{0.5cm}p{0.4cm}p{1cm}p{4cm}p{0.4cm}p{0.6cm}p{1.3cm}p{0.5cm}p{0.4cm}p{0.6cm}R{0.9cm}p{0.5cm}}
    
        \hline
          
        J1429+3230C & 217.4756 & 32.5127 & 6.1 & I/II & 1.17 & 0.26 & p & 1,2,3,4 & SDSS J142954.06+323045.8 & Qc & 3.0 & 22.67 r & 11.6 & 1.6 & 25.8 & ~~0.81 & 0.12 \\   
        J1430+3519$^{2,7}$ & 217.5410 & 35.3329 & 4.0 & II & 0.629 &  & s & 3 & SDSS J143009.83+351958.2 & G & 1.7 & 20.76 r & 52.0 & 5.3 & 25.9 & 1.74 & 0.23\\ 
        J1430+3322C & 217.6154 & 33.3804 & 3.7 & I & 0.512 &  & s & 3 & SDSS J143027.69+332249.4 & G & 1.4 & 19.52 r & 14.6 & 1.5 & 25.1 & $>$0.11 & \\ 
        J1431+3345$^{1,2,5}$ & 217.7644 & 33.7616 & 3.6 & I/II & 0.238 &  & s & 3 & 2MASX J14310340+3345414 & G & 0.8 & 16.77 r & 166 & 17 & 25.4 & 0.67 & 0.04 \\
        J1431+3535C & 217.8483 & 35.5976 & 2.6 & II & 1.04 & 0.18 & p & 1,2,4 & DESI J217.8484+35.5975 & G & 1.3 & 23.01 r & 24.0* & 2.7 & 26.0 & ~ \\ 
        J1431+3440 & 217.8485 & 34.6681 & 2.2 & II & 0.79 & 0.09 & p & 1,2,3,4 & SDSS J143123.56+344005.0 & G & 1.0 & 21.68 r & 8.2 & 0.9 & 25.3 & ~ \\ 
        J1431+3427$^{1,2}$ & 217.8639 & 34.4506 & 3.0 & II & 0.44 & 0.15 & p & 1,3 & SDSS J143127.33+342702.3 & G & 1.0 & 20.82 r & 116 & 11 & 25.9 & 0.88 & 0.04 \\ 
        J1432+3411$^{1,2}$ & 218.0479 & 34.1926 & 2.4 & II & 1.04 & 0.15 & p & 1,2,3,4,5 & SDSS J143211.50+341133.2 & G & 1.2 & 22.68 r & 65.7 & 6.6 & 26.5 & 1.03 & 0.05\\ 
        J1432+3328$^6$ & 218.1046 & 33.4770 & 1.8 & II & 1.30 & 0.04 & p & 1,2,4,5 & DESI J143223.46+332852.5 & Qc & 0.9 & 21.84 r & 114 & 11 & 26.9 & 1.13 & 0.04 \\ 
        J1432+3545$^2$ & 218.1413 & 35.7612 & 5.0 & I & 0.185 & ~ & s & 3 & 2MASX J14323389+3545397 & G & 0.9 & 16.71 r & 84.0 & 8.5 & 24.9 & 0.77 & 0.04\\ 
        J1432+3220$^{1,2}$ & 218.1777 & 32.3459 & 1.6 & II & 0.81 & 0.08 & p & 1,2,4 & SDSS J143242.64+322045.4 & G & 0.7 & 22.30 r & 107 & 11 & 26.4 & ~ \\ 
        J1432+3154C & 218.2283 & 31.9030 & 3.3 & II & 0.88 & 0.15 & p & 1 & DESI J218.2283+31.9030 & G & 1.6 & 23.15 r & 5.2 & 1.7 & 25.2 & 0.68 & 0.15 \\ 
        J1432+3647 & 218.2430 & 36.7894 & 1.5 & II & 2.836 & ~ & s & 3 & SDSS J143258.30+364721.7 & Q & 0.7 & 22.20 r & 13.1 & 1.6 & 26.8 & 0.52 & 0.25 \\ 
        J1433+3220 & 218.2569 & 32.3435 & 5.9 & II & 0.552 & ~ & s & 3 & SDSS J143301.66+322036.7 & G & 2.3 & 20.40 r & 12.3 & 2.5 & 25.12 & ~ \\ 
        J1433+3328 & 218.3156 & 33.4830 & 1.4 & I/II & 1.609 & ~ & s & 8 & SDSS J143315.74+332858.6 & Q & 0.7 & 22.76 r & 28.4 & 2.9 & 26.6 & 0.68 & 0.05 \\ 
        J1433+3450$^{2,3}$ & 218.3258 & 34.8485 & 1.7 & II & 1.00 & 0.14 & p & 1,2,4,5 & DESI J218.3257+34.8485 & G & 0.8 & 22.79 r & 1628 & 163 & 27.8 & 1.00 & 0.04\\ 
        J1433+3145C & 218.3367 & 31.7665 & 2.4 & II & 0.69 & 0.06 & p & 1,3,4 & SDSS J143320.81+314559.2 & G & 1.0 & 22.13 r & 5.1 & 1.4 & 24.9 & 0.84 & 0.15 \\ 
        J1434+3428$^{1,2}$ & 218.6212 & 34.4668 & 2.6 & II & 0.96 & 0.04 & p & 1,2,3,4 & SDSS J143429.09+342800.4 & G & 1.2 & 22.71 r & 145 & 14 & 26.7 & 0.96 & 0.04 \\ 
        J1434+3214C & 218.6639 & 32.2407 & 2.7 & I & 1.5 & ~ & e & - & CWISE J143439.34+321426.6 & ? & 1.4 & 17.83 W2 & 3.7 & 1.2 & 25.6 & 0.40 & 0.18 \\ 
        J1434+3648 & 218.6889 & 36.8099 & 2.3 & I & 0.429 & ~ & s & 3 & SDSS J143445.33+364835.5 & G & 0.8 & 19.27 r & 10.6 & 1.4 & 24.8 & ~ \\ 
        J1434+3328C & 218.6890 & 33.4724 & 4.7 & II & 0.198 & ~ & s & 3 & 2MASX J14344537+3328211 & Q & 0.9 & 16.72 r & 54.8 & 5.7 & 24.8 & ~ \\ 
        J1435+3404 & 218.8812 & 34.0780 & 2.1 & II & 0.546 & ~ & s & 3 & SDSS J143531.49+340440.9 & G & 0.8 & 21.82 r & 5.6* & 0.8 & 24.8 & ~ \\ 
        J1435+3547 & 218.9370 & 35.7978 & 1.7 & II & 1.63 & 0.27 & p & 1,2,3,4 & SDSS J143544.87+354752.2 & G & 0.9 & 22.72 r & 30.1 & 3.1 & 26.6 & 1.43 & 0.19  \\ 
        J1436+3416 & 219.2216 & 34.2807 & 3.7 & II & 0.25 & 0.15 & p & 1,2,3,4 & SDSS J143653.19+341650.5 & G & 0.9 & 21.67 r & 69.8 & 7.1 & 25.1 & 0.95 & 0.05 \\ 
        J1437+3650 & 219.3635 & 36.8387 & 3.3 & II & 0.94 & 0.09 & p & 1,3,4 & SDSS J143727.24+365019.3 & G & 1.6 & 23.31 r & 8.0 & 1.6 & 25.5 & 0.67 & 0.15 \\ 
        J1437+3233 & 219.4135 & 32.5566 & 1.5 & I & 1.02 & 0.21 & p & 1,3,4 & SDSS J143739.23+323323.8 & G & 0.7 & 23.46 r & 6.3 & 0.8 & 25.4 & 0.75 & 0.13 \\ 
        J1438+3445C & 219.5403 & 34.7591 & 3.2 & II & 0.58 & 0.05 & p & 1,2,3,4,7 & SDSS J143809.66+344532.8 & G & 1.3 & 21.11 r & 9.5 & 1.5 & 25.1 & ~ \\ 
        J1438+3526 & 219.5529 & 35.4368 & 1.6 & II & 0.86 & 0.08 & p & 1,3 & SDSS J143812.69+352612.6 & G & 0.8 & 22.32 r & 11.0 & 1.2 & 25.5 & 0.90 & 0.11 \\ 
        J1438+3355 & 219.5877 & 33.9290 & 7.5 & II & 1.80 & 0.17 & p & 2,5 & SDSS J143821.02+335544.4 & Qc & 3.8 & 15.77 W2 & 21.5 & 2.7 & 26.5 & $>$1.17 \\ 
        J1438+3539 & 219.6260 & 35.6545 & 7.1 & II & 0.262 & ~ & s & 3,10 & SDSS J143830.23+353916.1 & Q & 1.7 & 18.36 r & 17.9 & 2.3 & 24.5 & 0.65 & 0.05 \\ 
        J1439+3251C & 219.7544 & 32.8586 & 3.0 & II & 0.57 & 0.11 & p & 1,3,4,6,9 & SDSS J143901.04+325131.0 & G & 1.2 & 21.41 r & 4.0 & 1.2 & 24.7 & 0.40 & 0.41 \\
        J1439+3254$^{1,2}$ & 219.8133 & 32.9137 & 1.7 & II & 0.80 & 0.04 & p & 1,3,4 & SDSS J143915.18+325449.4 & G & 0.8 & 21.35 r & 228 & 23 & 26.8 & 0.88 & 0.04 \\ 
        J1439+3221C & 219.9937 & 32.3575 & 4.9 & II & 0.50 & 0.02 & p & 1,3,4 & SDSS J143958.48+322126.9 & G & 1.8 & 20.54 r & 23.2* & 3.1 & 25.3 & $>$0.77 \\ 
        J1440+3211 & 220.0520 & 32.1865 & 7.5 & II & 0.49 & 0.12 & p & 1,3,4 & SDSS J144012.47+321111.3 & G & 2.7 & 20.48 r & 149 & 15 & 26.1 & 0.61 & 0.15 \\ 
        J1440+3348 & 220.1655 & 33.8009 & 2.0 & I/II & 0.542 & ~ & s & 3 & SDSS J144039.70+334803.2 & G & 0.8 & 20.43 r & 17.4 & 1.9 & 25.2 & ~ \\ 
        J1442+3457 & 220.5530 & 34.9617 & 3.0 & II & 0.532 & ~ & s & 3 & SDSS J144212.71+345742.2 & G & 1.1 & 20.17 r & 9.7* & 1.2 & 25.0 & 0.56 & 0.14 \\ 
        J1442+3605 & 220.6305 & 36.0897 & 1.8 & II & 1.10 & 0.15 & p & 1 & DESI J220.6304+36.0893 & G & 0.9 & 23.17 r & 457 & 46 & 27.4 & ~ \\ 
        J1442+3243$^5$ & 220.6456 & 32.7242 & 2.9 & II & 0.48 & 0.20 & p & 1,3,7 & SDSS J144234.93+324326.9 & Qc & 1.1 & 19.02 r & 344 & 34 & 26.4 & ~ \\ 
        J1442+3242 & 220.6844 & 32.7163 & 1.9 & II & 0.80 & 0.05 & p & 1,3 & SDSS J144244.26+324258.6 & G & 0.8 & 22.30 r & 166 & 17 & 26.6 & ~ \\ 
        J1444+3445C & 221.0615 & 34.7596 & 3.3 & II & 0.259 & ~ & s & 3 & SDSS J144414.75+344534.5 & G & 0.8 & 17.69 r & 16.5* & 1.9 & 24.5 & ~ \\ 
        J1444+3348 & 221.1075 & 33.8054 & 1.8 & ? & 0.91 & 0.07 & p & 1,3,4 & SDSS J144425.80+334819.3 & G & 0.9 & 22.69 r & 15.0 & 2.2 & 25.7 & ~ \\ 
        J1444+3444 & 221.1722 & 34.7425 & 1.7 & II & 0.77 & 0.03 & p & 1,3,4 & SDSS J144441.33+344433.0 & G & 0.8 & 21.71 r & 39.6 & 4.3 & 26.0 & 0.64 & 0.05 \\ \hline
    \end{tabular}
    
\end{sidewaystable*}

\begin{figure*}
        \centering
        \includegraphics[width= 1\textwidth]{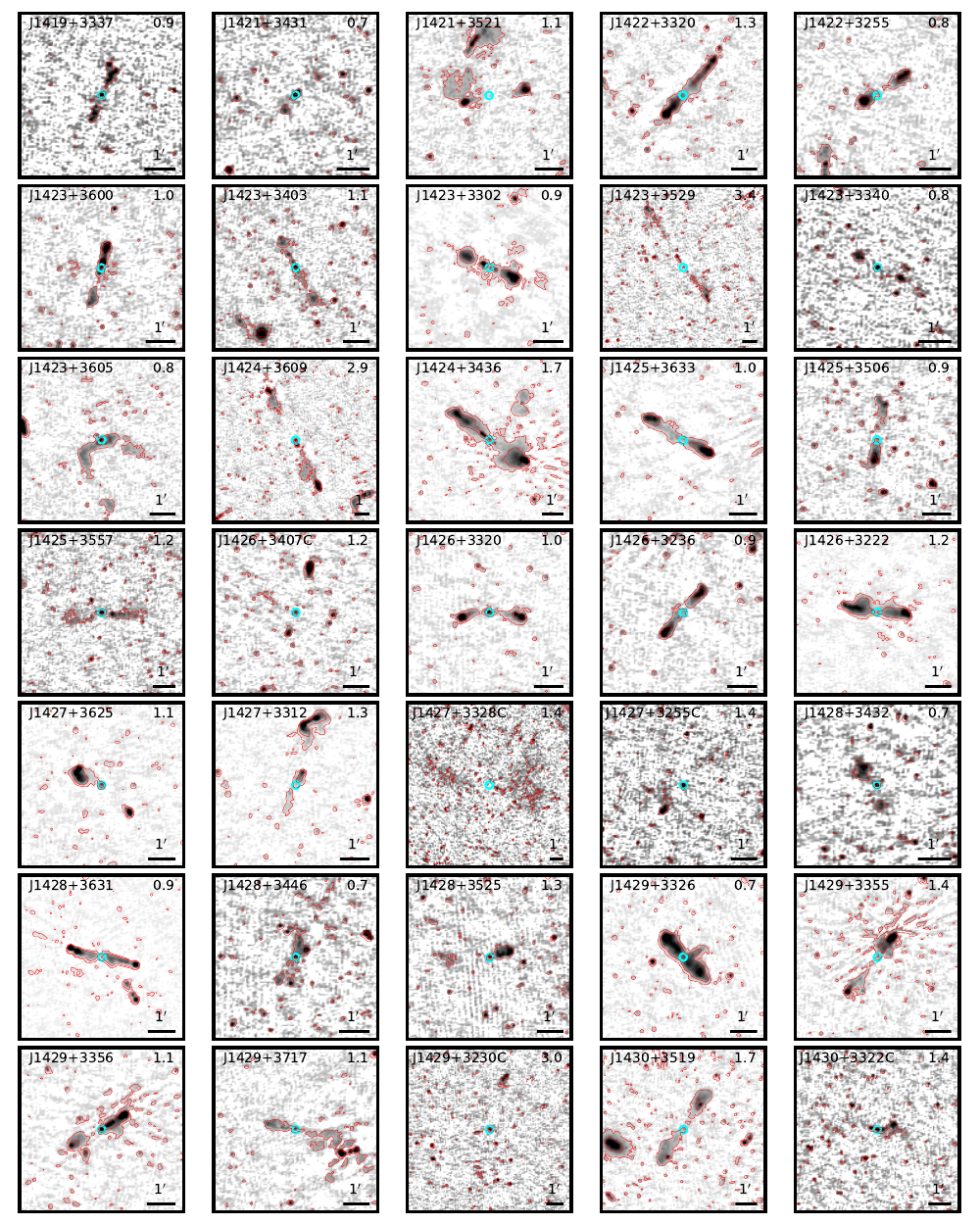}
        \caption{Cutouts of the BLDF image around our GRGs at 150 MHz with 3 and 24-$\sigma_{rms}$ red contours superimposed.  The resolution of the images is 6$ \rm "$. The cyan circle identifies the position of the host galaxy. The bar in the bottom-right corner represents an angular size of 1$ \rm '$, while the name and the largest linear size (in Mpc) of the GRGs are reported in the top-left and top-right corners, respectively.}
        \label{fig:GRG_images}
  \end{figure*}
  
  \begin{figure*}
        \centering
        \includegraphics[width= 1\textwidth]{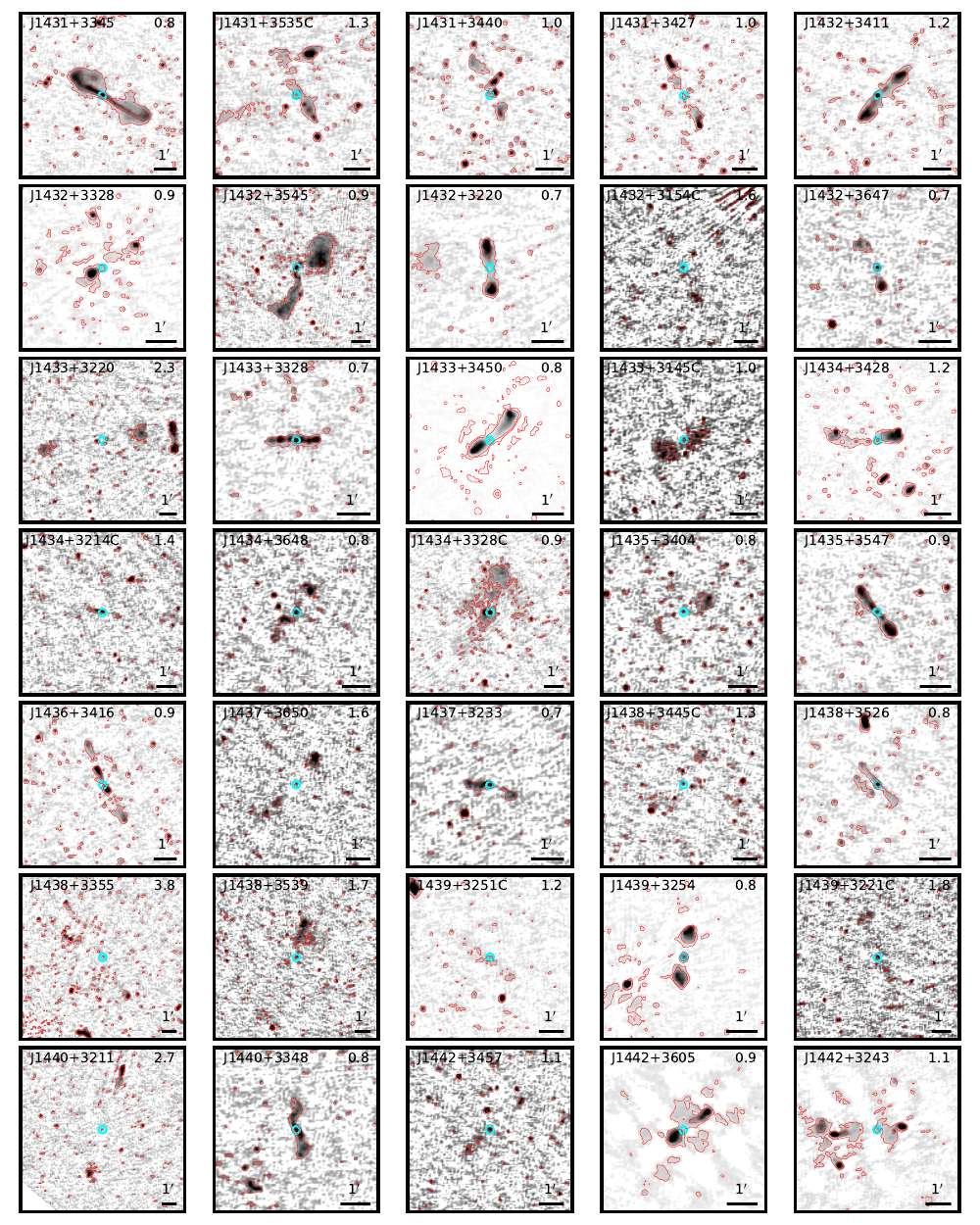}
  \end{figure*}
  
  \begin{figure*}
        \centering
        \includegraphics[width= 1\textwidth]{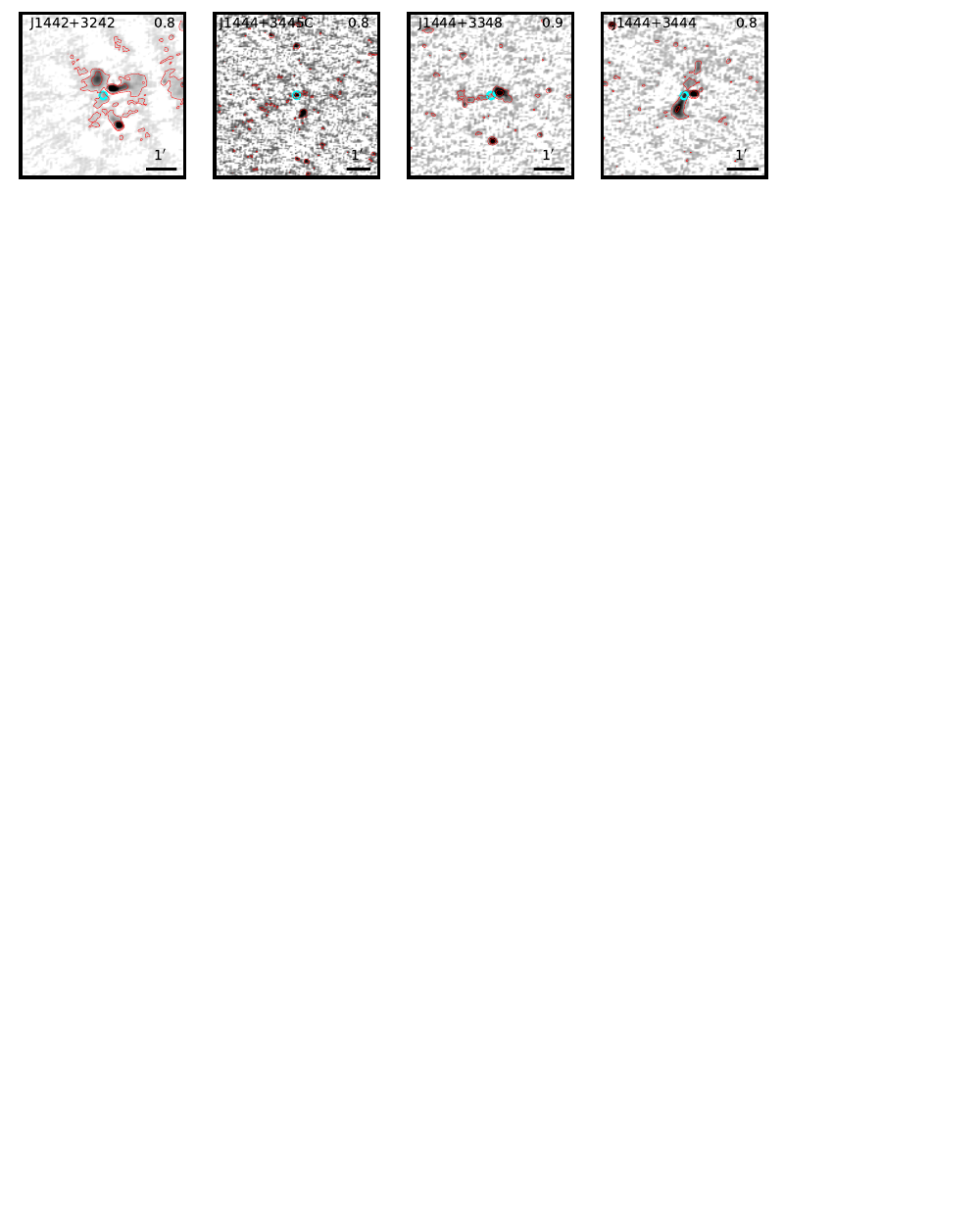}
  \end{figure*}

\bsp	
\label{lastpage}
\end{document}